\documentclass[aps,article,twocolumn,longbibliography,citeautoscript,footinbib,superscriptaddress]{revtex4-2} \synctex=1 
\usepackage{amsmath,amssymb,bm,bbm} 
\usepackage{amsthm}
\usepackage{physics}
\usepackage{graphicx}  
\usepackage{color} 
\usepackage[dvipsnames]{xcolor}
\usepackage[colorlinks=true]{hyperref}  
\usepackage[section]{placeins}
\usepackage[mathscr]{euscript}
\usepackage[toc,page]{appendix}
\hypersetup{
    unicode=false,          
    pdftoolbar=true,        
    pdfmenubar=true,        
    pdffitwindow=false,     
    pdfstartview={FitH},    
    pdftitle={Maximal quantum chaos of the critical Fermi surface},    
    pdfauthor={Maria Tikhanovskaya, Subir Sachdev, and Aavishkar A. Patel},     
    pdfsubject={},   
    pdfcreator={},   
    pdfproducer={}, 
    pdfkeywords={} {} {}, 
    pdfnewwindow=true,      
    colorlinks=true,       
    linkcolor=magenta, 
    citecolor=blue,        
    filecolor=magenta,      
    urlcolor=blue           
} 

\usepackage{pdfpages}
\usepackage{pgffor}
\makeatletter
\AtBeginDocument{\let\LS@rot\@undefined}
\makeatother

\usepackage{tikz,fp}
\usepackage{tikz-cd}
\usetikzlibrary{arrows}
\usetikzlibrary{intersections}
\usetikzlibrary{shapes.geometric}
\usetikzlibrary{decorations.pathmorphing, patterns,shapes,fixedpointarithmetic}
\usetikzlibrary{decorations.markings}

\pgfdeclarepatternformonly{south west lines}{\pgfqpoint{-0pt}{-0pt}}{\pgfqpoint{3pt}{3pt}}{\pgfqpoint{3pt}{3pt}}{
        \pgfsetlinewidth{0.4pt}
        \pgfpathmoveto{\pgfqpoint{0pt}{0pt}}
        \pgfpathlineto{\pgfqpoint{3pt}{3pt}}
        \pgfpathmoveto{\pgfqpoint{2.8pt}{-.2pt}}
        \pgfpathlineto{\pgfqpoint{3.2pt}{.2pt}}
        \pgfpathmoveto{\pgfqpoint{-.2pt}{2.8pt}}
        \pgfpathlineto{\pgfqpoint{.2pt}{3.2pt}}
        \pgfusepath{stroke}}

\tikzset{
  mid arrow/.style={postaction={decorate,decoration={
        markings,
        mark=at position .575 with {\arrow{stealth}}
      }}},
  near arrow/.style={postaction={decorate,decoration={
        markings,
        mark=at position .275 with {\arrow{stealth}}
      }}},
  far arrow/.style={postaction={decorate,decoration={
        markings,
        mark=at position .800 with {\arrow{stealth}}
      }}},
  snake arrow/.style={fixed point arithmetic, decorate, decoration={snake,amplitude=2pt, segment length=11pt},postaction={decoration={markings,mark=at position 0.625 with {\arrow{stealth}}},decorate}},
}

\usepackage{amsfonts}
\usepackage{upgreek}
\usepackage{slashed}
\usepackage{latexsym}
\usepackage{mathrsfs}
\usepackage{tikz-feynman}
\usepackage[usestackEOL]{stackengine}
\newcommand{\beq}{\begin{eqnarray}}
\newcommand{\eeq}{\end{eqnarray}}

\newcommand{\bx}{{\bm x}}

\newcommand{\bp}{{\bm p}}
\newcommand{\bk}{{\bm k}}
\newcommand{\bkp}{{{\bm k}'}}
\newcommand{\sgn}{\text{sgn}}

\begin{document}
\title{Maximal quantum chaos of the critical Fermi surface}

\author{Maria Tikhanovskaya}
\affiliation{Department of Physics, Harvard University, Cambridge MA 02138, USA}

\author{Subir Sachdev}
\affiliation{Department of Physics, Harvard University, Cambridge MA 02138, USA}
\affiliation{School of Natural Sciences, Institute for Advanced Study, Princeton, NJ-08540, USA}

\author{Aavishkar A. Patel}
\affiliation{Department of Physics, University of California Berkeley, Berkeley CA 94720, USA}

\date{\today}

\begin{abstract}
We investigate the many-body quantum chaos of non-Fermi liquid states with Fermi surfaces in two spatial dimensions by computing their out-of-time-order correlation functions. Using a recently proposed large $N$ theory for the critical Fermi surface, and the ladder identity of Gu and Kitaev, we show that the chaos Lyapunov exponent takes the maximal value of $2 \pi k_B T/\hbar$, where $T$ is the absolute temperature. We also examine a phenomenological model in which the chaos exponent becomes smaller than the maximal value precisely when quasiparticles are restored.
\end{abstract}

\maketitle

The study of relaxational and thermalization phenomena in quantum many body systems has long relied on the quasiparticle decomposition of many body states, and the collisions of quasiparticles described by the Boltzmann equation and its generalizations. However, this powerful method is not reliable when we address similar phenomena in non-Fermi liquids without any quasiparticle excitations. General arguments have been presented that such dissipative phenomena cannot occur at a rate which is parametrically larger than $k_B T/\hbar$ as the absolute temperature $T \rightarrow 0$ (so such a rate cannot vanish as $ T^a$, with an exponent $a<1$), and systems without quasiparticles have a rate of order $k_B T/\hbar$ \cite{Hartnoll:2021ydi,Chowdhury:2021qpy,qptbook,bruin,Zaanen,Gael21}.

New insights into such issues have emerged from recent advances in the study of many-body quantum chaos and out-of-time-order correlators (OTOCs), for which the bounds on dissipative rates can be made precise. Inspired by holographic connections to the quantum dynamics of black holes, Maldacena, Shenker and Stanford \cite{MSS16} established that the Lyapunov rate, $\lambda_L$, characterizing the temporal growth of the OTOC must be smaller than $2 \pi k_B T /\hbar$. We can expect that any system which is close to this bound as $T \rightarrow 0$ cannot have a quasiparticle description, and this conclusion is supported by computations on the Sachdev-Ye-Kitaev (SYK) model \cite{Maldacena_syk,kitaev2018}. Although difficult to measure in experiments, OTOCs 
have therefore emerged as an alternative to the Boltzmann equation, and are a valuable diagnostic of the physics of non-quasiparticle systems.

In this paper, we address the OTOC of a class of non-Fermi liquids most relevant to correlated electron systems \cite{SungSik18}. We consider a Fermi surface coupled to a U(1) gauge field in two spatial dimensions, but our theory applies also to Fermi surfaces to
coupled to other critical bosons, as are realized near symmetry-breaking quantum phase transitions in metals with a zero momentum order parameter. The OTOC of such a system was addressed in previous work \cite{patel2017quantum} in an uncontrolled analysis: it was found that $\lambda_L = \alpha k_B T/\hbar$ as $T \rightarrow 0$ with the constant $\alpha < 2 \pi$. The present paper will present new results on this model which build on two recent developments:\\
({\it i\/}) Gu and Kitaev (GK) \cite{gu2019} have shed new light on the structure of OTOCs in spatially extended systems. They established a ladder identity which shows that there is an additional contribution to the OTOC that arises from a pole at imaginary momentum, in the complex momentum plane. Provided the chaos butterfly velocity, $v_B$, is large enough, the pole contribution dominates at large spatial distances, and the resulting growth of the OTOC with time has exponent $\lambda_L$ exactly equal to $2 \pi k_B T/\hbar$. \\
({\it ii\/}) A systematic approach to the study of the two-dimensional non-Fermi liquid state has been proposed \cite{esterlis2021,Aldape2020}. This approach obtains the non-Fermi liquid as the large $N$ saddle-point of a path integral over bilocal Green's functions and self energies. The new idea here is to study an ensemble of theories with different random couplings (but without spatial randomness), under the hypothesis that all of them flow to the same universal fixed point theory at low energies. Such a large $N$ saddle-point is ideally suited to develop a systematic computation of the OTOC, along the lines of computations on the SYK model. 

In our large $N$ analysis of the two-dimensional non-Fermi liquid, we 
find that the butterfly velocity does indeed satisfy the needed inequality of GK.
This leads to our main result: that the Lyapunov rate of this systems equals the maximal value of $2 \pi k_B T /\hbar$.

\begin{figure}
\center{\includegraphics[width=2in]{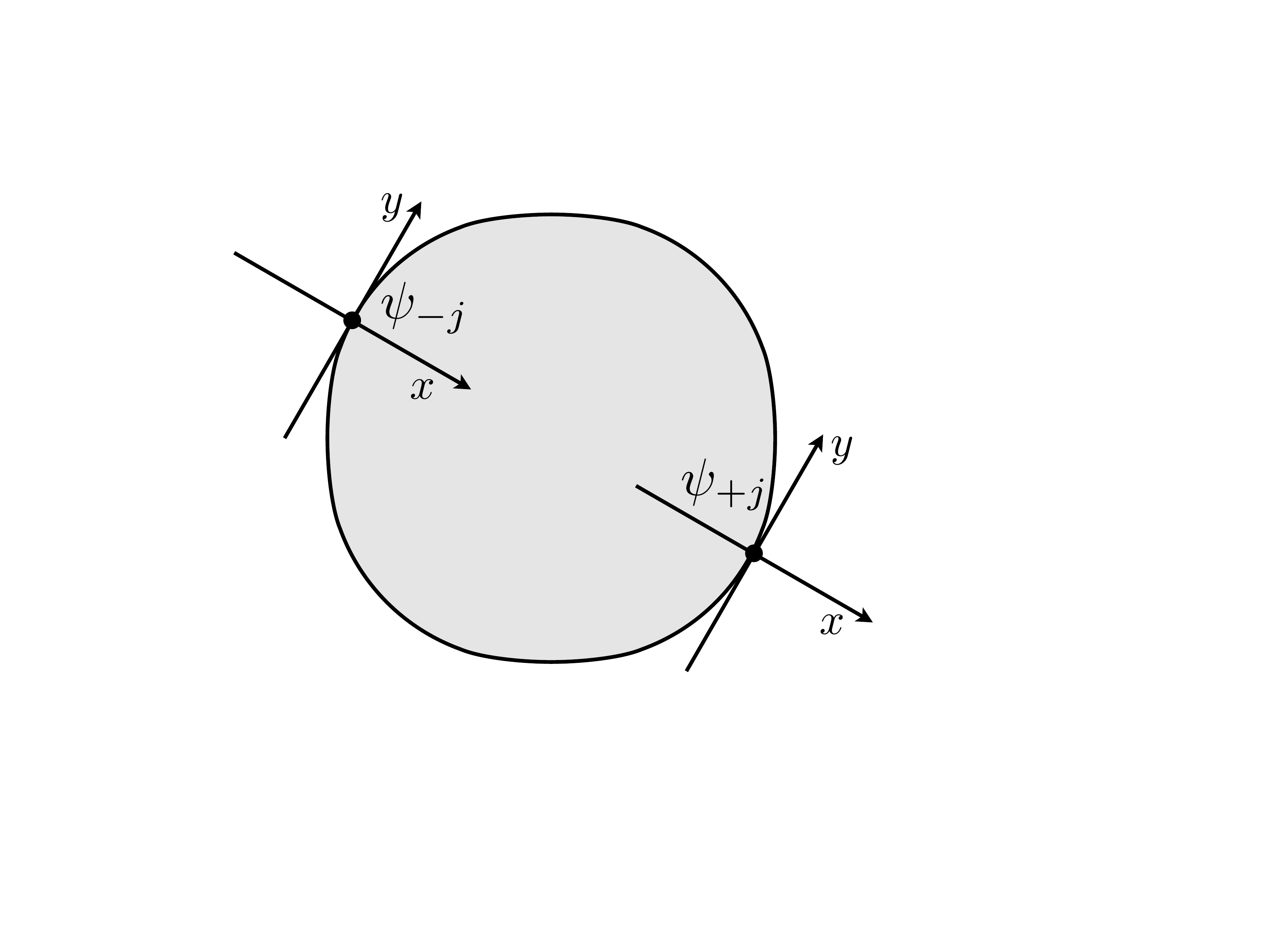}}
\caption{Antipodal patches on the Fermi surface with fermions $\psi_{\pm j}$, and the coordinate system.}
\label{fig:patch}
\end{figure}

{\it The model.} Our results are obtained within the `patch' theory of the non-Fermi liquid \cite{Lee:2009epi,metlitski1}, which describes the low energy properties of the Fermi surface without quasiparticles. Each point on the Fermi surface is characterized by a Fermi velocity $v_F$, and Fermi surface curvature $\kappa/v_F$. We introduce fermion fields $\psi_{\pm j}$ ($j = 1 \ldots N$) defined in patches near antipodal points, using the co-ordinate system shown in Fig.~\ref{fig:patch}.
The dispersion of the fermions in this co-ordinate system is $\varepsilon_{\bm k} =  \pm v_F k_x + (\kappa/2) k_y^2$, and we will henceforth use length scales in which $v_F=1$ and $\kappa=2$. These fermions are coupled to a boson $\phi_l$ ($l = 1 \ldots M$), which is the transverse component of the gauge field, or a symmetry breaking order parameter. The universal properties of the critical Fermi surface in this patch theory are then described by the 
2+1 dimensional Lagrangian density 
\begin{align}
\mathcal{L} &= \sum_{j=1}^N \psi_{+j}^\dagger(\partial_\tau - i\partial_x -\partial_y^2)\psi_{+j} + \sum_{l=1}^M \sum_{ij=1}^N \frac{g_{ijl}}{N}\psi_{+i}^\dagger \psi_{+j} \phi_l \nonumber \\
&+ \sum_{j=1}^N \psi_{-j}^\dagger(\partial_\tau + i\partial_x -\partial_y^2)\psi_{-j} +  s \sum_{l=1}^M \sum_{ij=1}^N \frac{g_{ijl}}{N}\psi_{-i}^\dagger \psi_{-j} \phi_l \nonumber \\
&~~~~~~~~~~~~~~~~~~+ \frac{1}{2} \sum_{i=1}^M (\partial_y\phi_i)^2 \,.
\label{otocL}
\end{align}
Here the sign $s=+1$ for the nematic order parameter, $s=-1$ for the gauge field; the sign of $s$ will not be important for any results here. 
The large $N$ limit \cite{esterlis2021} is taken at fixed $M/N$, and for an ensemble of theories with spatially uniform (but flavor random) Yukawa couplings $g_{ijl}$ which have zero mean and root mean square value $g$ ($\overline{|g_{ijl}|^2} = g^2$). The scaling limit of the boson ($D(k)$) and fermion 
($G(k)$) Green's functions can be computed exactly in the large $N$ limit ($k = (\mathbf{k},k_0) = (k_x, k_y, k_0)$, where $k_0$ is an imaginary Matsubara frequency)
\beq
D(k) &=& \frac{|k_y|}{|k_y|^3 +c_b |k_0| + m^2}, \nonumber \\
\left[G(k)\right]^{-1} &=& k_x + k_y^2  - i\mu(T)\sgn(k_0) \\
&-& ic_f\sgn(k_0)T^{2/3}H_{1/3}\left(\frac{|k_0|-\pi T\sgn(k_0)}{2\pi T}\right) \,, \nonumber
\eeq
where $H_{1/3} (z) =\zeta(1/3) - \zeta(1/3,z+1) $ the analytically continued harmonic number function of order 1/3, and $c_f$ and $c_b$ are coupling dependent constants:  
\beq
c_f = \frac{M}{N} \frac{2^{4/3} g^{4/3}}{3^{3/2}} \,, \quad c_b = \frac{g^2}{4\pi} \,. 
\eeq
We have also introduced a finite but small mass $m^2$ in the boson Green's function as an infrared regulator, and $\mu(T) = g^2T/(3\sqrt{3}m^{2/3})$. We will eventually take the $m \rightarrow 0$ limit, and obtain a finite answer for the OTOC.
To solve for the OTOC we use retarded and Wightman Green's functions, the forms of which we discuss in the Supplementary Information.  

{\it The OTOC.}
We will be interested in the OTOC contianed within the squared anticommutator of fermionic operators 
\beq
\mathcal{C}_{\bm x} (t,0) &=& \frac1{N^2} \theta(t)\sum_{i,j=1}^N \text{Tr} \Bigl[ e^{-\beta H/2} \{ \psi_i ({\bm x},t) , \psi_j^\dagger(0) \} \nonumber \\
&~&~\times e^{-\beta H/2} \{ \psi_i({\bm x},t) , \psi_j^\dagger(0) \}^\dagger \Bigr]. \label{otoc1}
\eeq
Note that we have dropped the $\pm$ index on the fermions, and all fermion operators in (\ref{otoc1}) belong to the same patch.
The function in (\ref{otoc1}) contains the out-of-time-ordered correlator $\langle\psi_i({\bm x},t)\psi_j^\dagger(0)\psi_i^\dagger({\bm x},t)\psi_j(0)\rangle$ (up to insertions of $e^{-\beta H/2}$), which in turn describes chaos in the system and has the exponential behavior $\sim e^{\lambda_L t}+\dots$, where $\lambda_L$ is the Lyapunov exponent. 
We are especially interested in the spatial structure of (\ref{otoc1}) in the long time limit at large $|{\bm x}|$. 
\begin{figure}
\center{\includegraphics[width=3in]{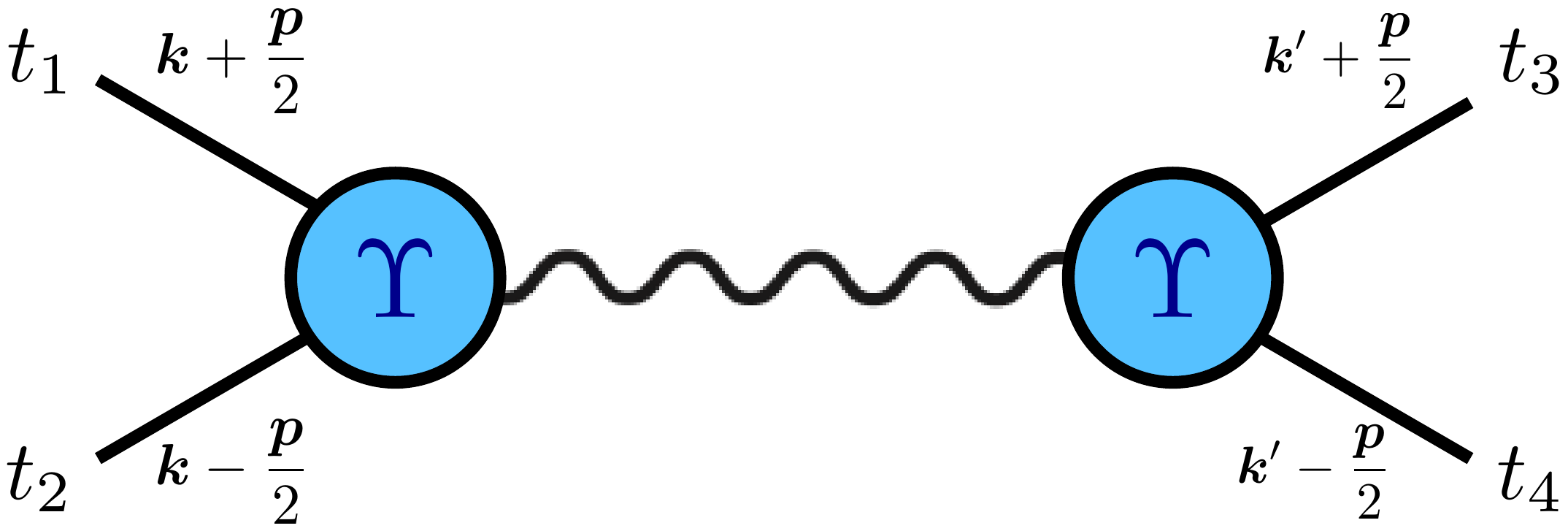}}
\caption{Diagram representing the ansatz \eqref{eq:ansatz1}. The wavy line represents the scramblon. The figure is adapted from \cite{gu2019} for the momentum-dependent case.}
\label{fig:ansatz} 
\end{figure}
After Fourier transforming the spatial arguments to momentum space, and considering 4 distinct times for the fermion operators (see Fig.~\ref{fig:ansatz}), Kitaev and Suh \cite{kitaev2018} argued that the early time OTOC could be written using a single mode ansatz involving the `scramblon' 
\beq\label{eq:ansatz1}
&& \text{OTOC}_{\bm p}(t_1,t_2,t_3,t_4;{\bm k},{\bm k}') \approx \nonumber \\
&&~~~~~\frac{e^{\lambda_L ({\bm p}) (t_1+t_2-t_3-t_4)/2}}{C({\bp})}  \Upsilon_{\bp}^R(t_{12},{\bk})\Upsilon_{p}^A(t_{34},{\bkp}) \label{otoc2}
\eeq
Here the $\Upsilon$'s are vertex functions which only modify the overall magnitude of the OTOC. It was later shown by GK that $C(\bp)$ is the important inverse propagator of the scramblon which leads to the exponential growth of chaos at rate $\lambda_L ({\bm p})$. As we review in the Supplementary Information, $C(\bp)$, has the important factor
\beq
C({\bp}) \sim \cos\frac{\lambda_L({\bp})}{4T}.
\eeq
which vanishes at the maximal chaos value $\lambda_L (\bp) = 2 \pi T$. The resulting pole in (\ref{otoc2}) will ultimately be responsible for the maximal chaos in the non-Fermi liquid.

Now we can transform back to position space and obtain
\beq
&& \text{OTOC}_{{\bm x}} (t_1, t_2, t_3, t_4) \sim  \frac{u ({\bm x}, t)}{N} \int_{\bk,\bkp}\!\!\! \Upsilon_{\bp}^R(t_{12},{\bk})\Upsilon_{p}^A(t_{34},{\bkp})\,,\nonumber\\\label{eq:OTOCx}
\eeq
where $t=(t_1+t_2-t_3-t_4)/2$ and 
\beq
u({\bm x},t) \sim \int_{\bp} \frac{e^{\lambda_L(\bp)t + i \bp \cdot {\bm x}}}{\cos( \lambda_L (\bp)/(4T))}. \label{otoc3}
\eeq
In the previous work \cite{patel2017quantum}, the chaos exponent was identified with $\lambda_L (0)$. GK performed a careful evaluation of the integral in (\ref{otoc3}) in one spatial dimension, and gave conditions under which it was dominated by the saddle point ($\lambda_L'(p = p_s)t + i x = 0$) or the pole ($\lambda_L (p = p_1) = 2 \pi T$). Both the saddle point and the pole appear for purely imaginary values of momenta, with $p = i |p|$. When $|p_s|>|p_1|$, GK showed that the pole dominates, leading to a region of spacetime in which maximal chaos occurs. Conversely, when $|p_s|<|p_1|$, the saddle point dominates, and there is no maximal chaos. 

At first sight, it is not clear whether this one-dimensional analysis can be extended to the anisotropic 2+1 dimensional non-Fermi liquid theory in (\ref{otocL}). However, the theory in (\ref{otocL}) has a `sliding symmetry' \cite{metlitski1}, which implies that $\lambda_L$ is a function only of $p_x + p_y^2$. This reduces the momentum integral in (\ref{otoc3}) to effectively a one-dimensional integral, and we can replace ${\bm p} \cdot {\bm r}$ by $p_x x$ and directly apply the results of GK.

\begin{figure}
\center{\includegraphics[width=3.3in]{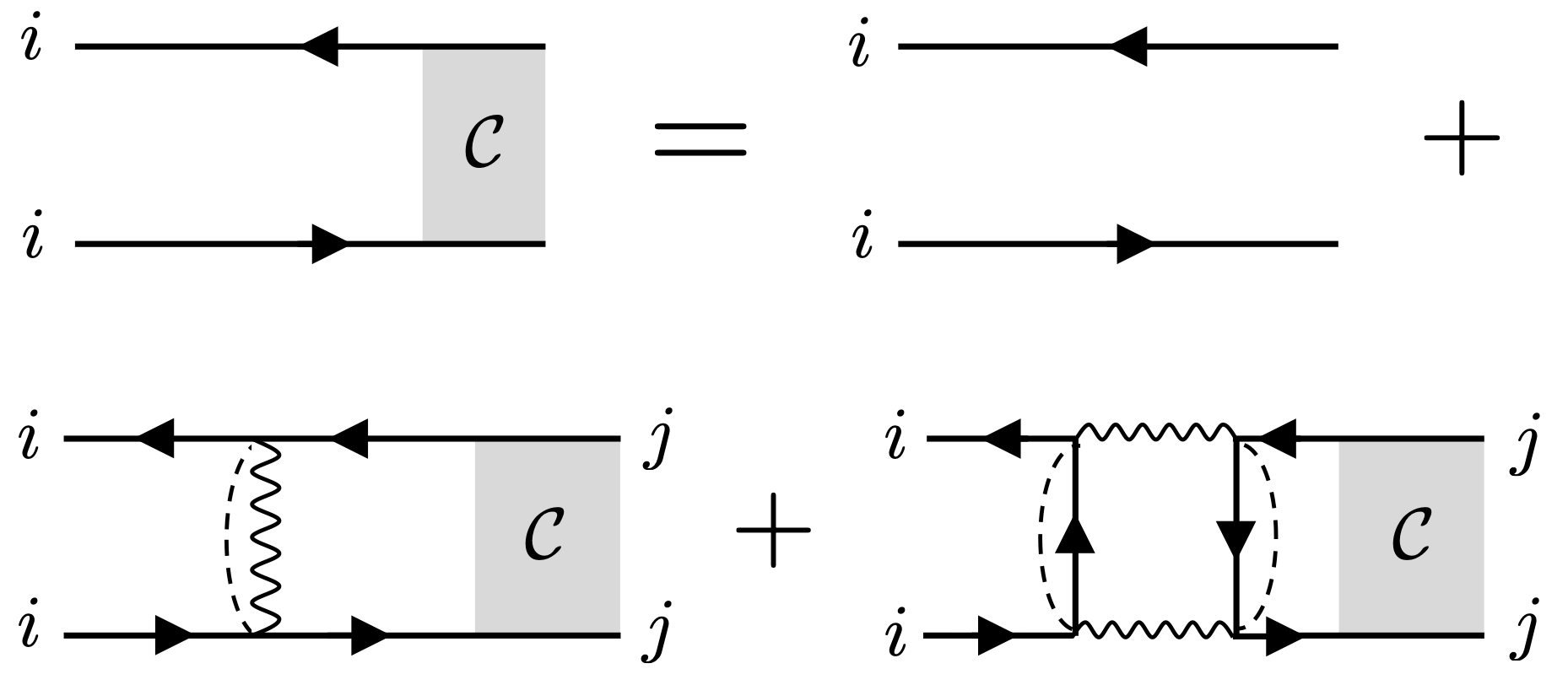}}
\caption{The Bethe-Salpeter equation for $\mathcal{C}(k_0,\omega,ip_x)$, which is exact at large $N$. Solid lines are fermion propagators, wavy lines are boson propagators and dashed lines are averaging over the flavor random couplings. The horizontal lines represent the retarded Green's functions and vertical lines are Wightman propagators.}
\label{fig:Bethe_Salpeter}
\end{figure}
{\it The Lyapunov exponent.}
The remaining missing ingredient in determining whether the saddle point or the pole dominates for the critical Fermi surface is a knowledge of $\lambda_L (p)$ for imaginary $p$. For this we need to solve the Bethe-Salpeter equation for the squared anticommutator $\mathcal{C}$ in Fig.~\ref{fig:Bethe_Salpeter}, with an imaginary external momentum. 
This leads to the following eigenvalue equation, extending the previously obtained equation \cite{patel2017quantum} to an imaginary external momentum $i p_x$ 
\begin{widetext}
\beq\nonumber
&&\left[c_f T^{2/3} \left(H_{1/3}\left(\frac{-ik_0-\pi T}{2\pi T}\right) + H_{1/3}\left(\frac{-i(\omega-k_0)-\pi T}{2\pi T}\right)\right) - p_x + 2\mu(T)\right] {\mathcal{C}}(k_0,\omega, ip_x) \\ \nonumber
&& = {g^2}\frac{M}{N}\int \frac{dk_0'dk'_y}{(2\pi)^2} \frac{c_b(k_0-k_0')|k_y'|}{ (|k_y'|^3 +m^2)^2 + c_b^2 (k_0-k_0')^2}\frac{{\mathcal{C}}(k_0',\omega,ip_x)}{\sinh\frac{k_0-k_0'}{2T}}\\
&&+\,\frac{g^{4/3}4\pi^{4/3}}{3\sqrt{3}}\frac{M}{N} 
\int\frac{dk_0' d k_{01}}{(2\pi)^2} \frac{(i k_{01} + (-i k_{01})^{2/3}(i(k_{01}-\omega)))}{ k_{01}(i(k_{01}-\omega))^{1/3}(2 k_{01}-\omega)}\,\frac{{\mathcal{C}}(k_0',\omega,ip_x)}{\cosh{\frac{k_0-k_{01}}{2T}}\,\cosh{\frac{k_0'-k_{01}}{2T}}}\label{eq:eigenPx}.
\eeq
\end{widetext}

Here, we are considering only the contributions of fermion propagators from a single patch on the Fermi surface \footnote{For the single patch theory, the values of $c_f,c_b$ are given by $c_f=(M/N)2^{5/3}g^{4/3}/3^{3/2}$ and $c_b = g^2/(8\pi)$.}: in the Supplementary Information, we show that the full equation that considers couplings between antipodal patches does not yield different results. Furthermore, we note that the factors of $M/N$ on the RHS of \eqref{eq:eigenPx} cancel with those in the definition of $c_f$ in the $m\rightarrow0$ limit, up to a rescaling of $p_x\rightarrow (N/M)p_x$. Therefore, considering $M\neq N$ will not affect any of our conclusions, and we will henceforth consider $M=N$ for simplicity.

\begin{figure}
\center{\includegraphics[width=2.95in]{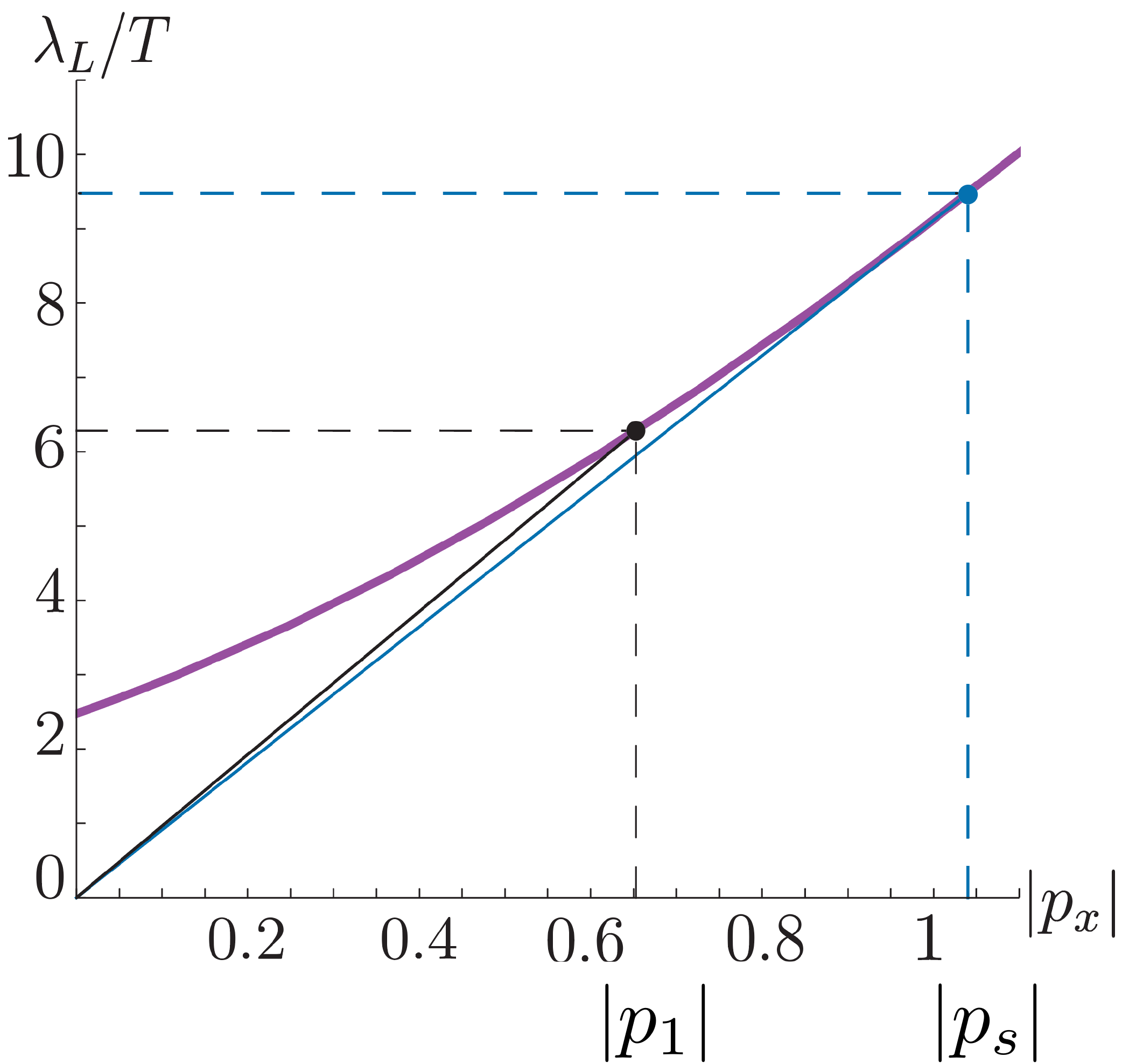}}
\caption{Plot of the Lyapunov exponent $\lambda_L/T$ as a function of external momentum $ip_x$ on imaginary axis. The absolute value of $p_x$ is presented in the units of $g^{4/3}T^{2/3}$. We find $|p_s|\approx 1.04\,g^{4/3} T^{2/3}$, $|p_1| \approx0.65\,g^{4/3} T^{2/3}$. Since $|p_s|>|p_1|$, the butterfly velocity is given by $v_B = 2\pi T/|p_1| \approx 9.67 g^{-4/3} T^{1/3}$ (slope of the black solid line). We also find the value of the velocity at the saddle point $v_s=9.01 \,g^{-4/3} T^{1/3}$ (slope of the blue solid line). As expected from the previous work \cite{patel2017quantum}, we find $\lambda_L(0) = 2.48\, T$.}
\label{fig:LambdaPx}
\end{figure}

{\it Maximal chaos.} 
Upon solving the eigenvalue equation \eqref{eq:eigenPx} we obtain the Lyapunov exponent as a function of the external imaginary momentum. From the numerical curve that is presented in Fig. \ref{fig:LambdaPx}, the value of the pole $|p_1|$ and the saddle point $|p_s|$ can be obtained. The pole contribution easily follows from the equation $\lambda_L(p=p_1) = 2\pi T$ whereas for the saddle point, one needs to consider an additional condition since the saddle point equation $\lambda'_L(p=p_s) t+ ix=0$ does not define the value of $|p_s|$. This condition follows from the fact that the ansatz \eqref{eq:OTOCx} is valid only in the regime where initial correlations and non-linear effects can be ignored, which is when $\text{OTOC}_{\bx} (t_1,t_2,t_3,t_4) \gg 1/N$. Therefore, the function can be estimated as $u(\bx,t)\sim1$. This gives the following condition on the saddle point value $\lambda_L(p=p_s)t+ip_s x =0$. Combining this equation with the saddle point equation, we can easily find the momentum $|p_s|$ form $\lambda'_L(p=p_s)= \lambda_L(p=p_s)/|p_s|$. 

As shown in Fig. \ref{fig:LambdaPx}, the momentum at which the pole appears $|p_1|$ is significantly smaller than the saddle point value $|p_s|$. Specifically, we find $|p_1| \approx 0.65 \,g^{4/3} T^{2/3}$ and $|p_s| \approx 1.04\,g^{4/3} T^{2/3} > |p_1|$, which confirms the dominance of the pole contribution according to GK. The chaos wavefront therefore travels with a butterfly velocity $v_B=2\pi T/|p_1|$ set by the pole contribution.   

We note that $\lambda_L(p_x)$ does not depend upon the coupling $g$, as $g$ can be removed by rescaling the external momentum $p_x \to p_x/g^{4/3}$. With no other dimensionful parameters in \eqref{eq:eigenPx}, this also implies that $\lambda_L$ is proportional to temperature.

{\it Phenomenological models.}
We extend our analysis to a general case of dynamic critical exponent $2<z\leq3$, in which quasiparticle excitations are still absent, and compute the Lyapunov exponents. For those theories, the boson Green's function has the form \cite{esterlis2021}
\beq\nonumber
D(k) = \frac{|k_y|}{|k_y|^{z}+c_b|k_0|+ m^2},
\eeq
and the fermion self energy scales as $\Sigma \sim i \mbox{sgn}(k_0) |k_0|^{2/z}$ ($z=3$ for the original theory discussed earlier). Since $|\Sigma|\gg|k_0|$, quasiparticle excitations are not well defined in terms of the fermion spectral function. 

The form of the eigenvalue equation for the OTOC changes slightly, and is discussed in the Supplementary Information. As we show in the Fig.~\ref{fig:LambdaPxZ}, for each of these theories the butterfly velocity is also given by $v_B = 2\pi T/|p_1|$, and the Lyapunov exponent is maximal. To show this, we first solve the eigenvalue equations up to the pole momentum $p_1$. We then compare the instantaneous slope at $p_1$, which we call $v_*$, and the velocity $v_1 = 2\pi T/|p_1|$. For each plot we obtain $v_1 > v_*$. Since each of the curves in Fig. \ref{fig:LambdaPxZ} is a positive, monotonically increasing, and convex function, this implies that $|p_s|>|p_1|$, and the pole contribution therefore dominates with maximal chaos just like in the $z=3$ case.  

\begin{figure}
\center{\includegraphics[width=3.2in]{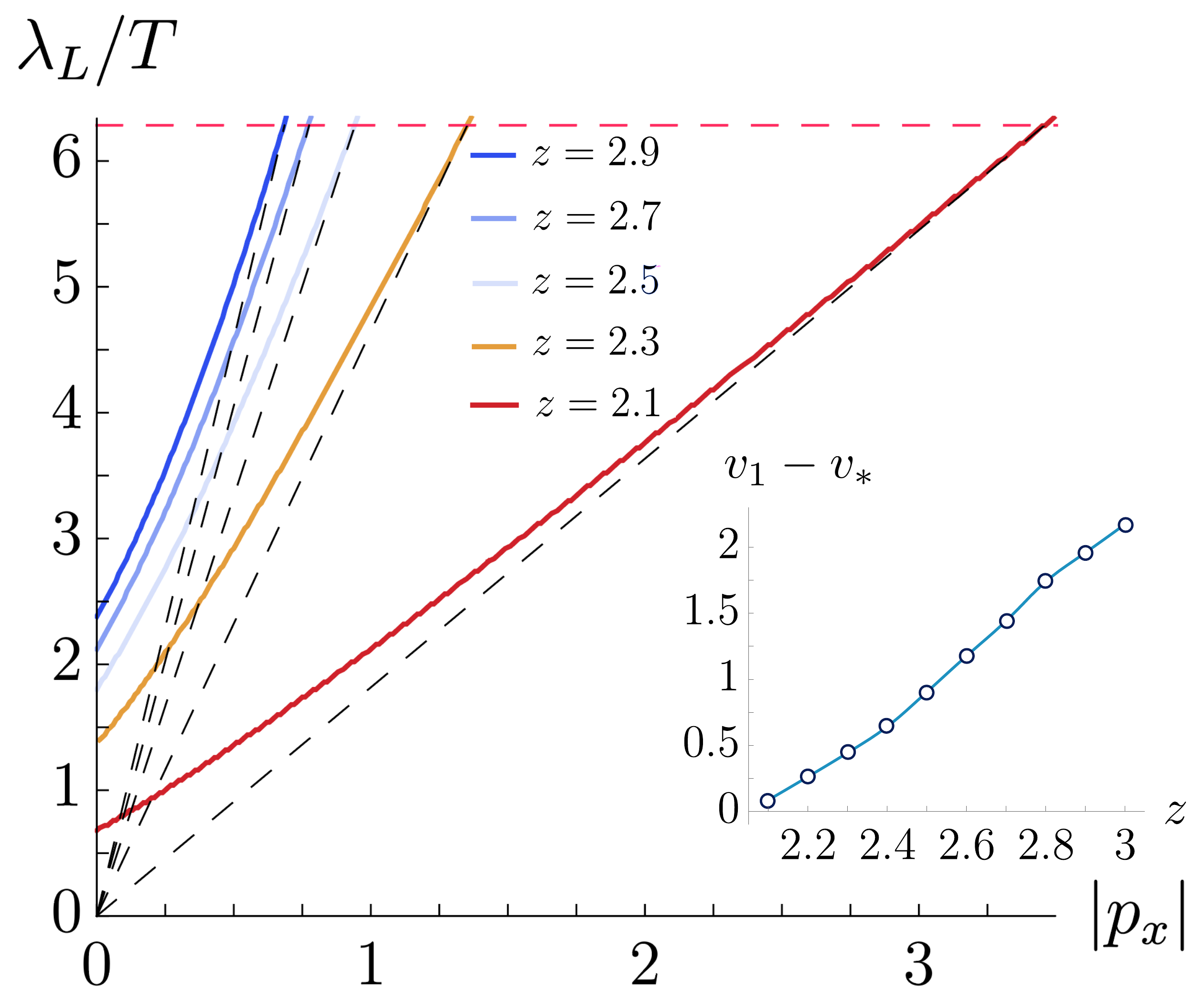}}\\
\caption{Main plot: resulting plots of the Lyapunov exponent $\lambda_L/T$ as a function of external momentum $p_x$ on the imaginary axis for different values of the dynamical critical exponent $2<z\leq 3$. The red dashed line shows where the Lyapunov exponent reaches $2\pi T$. Slopes of the black dashed lines are the butterfly velocities $v_B$. Inset: difference between slopes of the black lines on the main plot ({\it i.e.} $v_1=v_B$), and the instantaneous slopes at $p_x=p_1$ ({\it i.e.} $v_*$), for $2<z\leq3$. The difference is always positive, {\it i.e.} $\lambda_L(i|p_1|)/|p_1|>\lambda_L'(i|p_1|)$, which shows that the value of $|p_s|$, where $\lambda_L(i|p_s|)/|p_s|=\lambda_L'(i|p_s|)$, must be larger than $|p_1|$, and therefore each theory is maximally chaotic according to GK.}
\label{fig:LambdaPxZ}
\end{figure}

We can further find the behavior of the exponent in the case when $1<z<2$, in which quasiparticles are present. In this regime, $|\Sigma|\sim|k_0|^{2/z} \ll |k_0|$, and therefore the quasiparticle peak in the fermion spectral function is well defined. Similar to the discussion above, we can compute the Lyapunov exponent as a function of external momentum on imaginary axis and explicitly find $|p_s|$ and $|p_1|$. For a particular case of $z=3/2$ \footnote{The numerical values of $|p_s|$ and $|p_1|$ are provided for chosen parameters $g=0.5$, $\Lambda = 50$, $T=1$, and $z=3/2$. See Supplementary Information for more details.}, we find that the saddle point dominates as $|p_s|\approx4.32\,
$, which is smaller than $|p_1|\approx7.84$. The resulting butterfly velocity in this case is $v_B \approx 0.8\,v_F$, where $v_F=1$ is the Fermi velocity. This result is expected from a general point of view: for a free fermion theory the exponent is a simply linear function of the external momentum $\lambda_L(i|p_x|) = |p_x|$ leading to the saddle point contribution at $|p_s| = 0$. We therefore expect that for any theory with quasiparticles, the pole contribution is negligible compared to a saddle point contribution, and the maximal chaos is therefore no longer present.

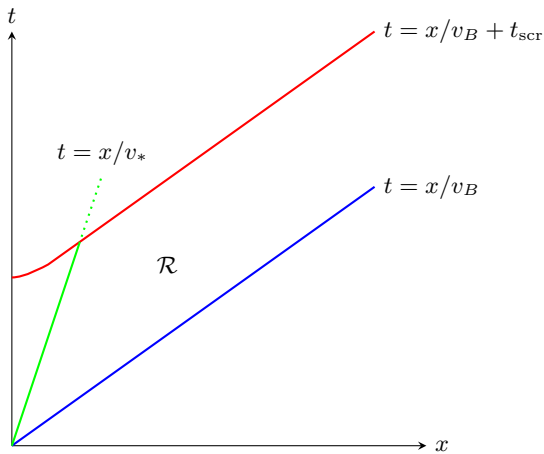
\begin{figure}
\center
\begin{tikzpicture}[scale=0.98, baseline={(current bounding box.center)}]
\draw[->,>=stealth] (0pt,0pt)--(160pt,0pt) node[right]{$x$};
\draw[->,>=stealth] (0pt,0pt) -- (0pt,160pt) node[above]{$t$};
\draw[thick, blue] (0pt,0pt) -- (140pt,100pt) node[black,  right]{$t=x/v_B$};
\draw[thick, red] (0pt,65pt).. controls (5pt, 65pt) and (12pt, 69pt) ..(14pt,70pt) -- (140pt,160pt) node[black,  right]{ $t=x/v_B+t_{\text{scr}}$};
\draw[thick, green] (0pt,0pt) -- (26pt,78pt);
\draw[thick, green, dotted] (26pt,78pt) -- (35pt,105pt) node[black, above]{$t=x/v_*$};
\node at (60pt,70pt){$\mathcal{R}$};
\end{tikzpicture}
\caption{
(Adapted from Ref. \cite{gu2019}) Maximal chaos is present in the spacetime region denoted by $\mathcal{R}$, which is bounded by the three solid lines. Here $v_* = \lambda_L'(i|p_1|)$, and $t_{\mathrm{scr}}\sim \ln N$. As $z$ is reduced from $3$ to $2$, $v_*$ approaches $v_B=v_1$ from below (Fig.~\ref{fig:LambdaPxZ}), and the size of the maximally chaotic region $\mathcal{R}$ is therefore squeezed to zero by the falling slope of the green line as $z\rightarrow 2$. For $z<2$, there is consequently no maximally chaotic region.}
\label{fig:squeeze}
\end{figure}

{\it Discussion.} It is quite remarkable that the generic low energy theory of Fermi surfaces without quasiparticles in $2+1$ dimensions display maximal chaos in the large $N$ limit. Other spatially extended quantum many body models connected to the SYK model ({\it e.g.} \cite{Gu2017XLQ,Altman2021}), and certain conformal field theories  \cite{turiaci2016cft,haehl2018effective,haehl2019reparametrization}, have been shown to display maximal chaos, but none of them have spatially dependent Green's functions {\it and} live in more than one space dimension. Besides displaying maximal chaos in $2+1$ dimensions without exhibiting local criticality, it is also remarkable that the critical Fermi surface does so without the presence of conformal symmetry, which also sets it apart from the previously mentioned examples that have conformal symmetry.

We believe that the maximal chaos of the Fermi surface is linked to the local nature of the singular self energy of the fermion at large $N$ {\it i.e.\/} the self energy is frequency dependent, but independent of momentum, a feature the Fermi surface theory shares with the SYK model (along with the local frequency-only dependence of the fermion pairing vertex \cite{esterlis2021}). There are small contributions to the fermion anomalous dimension at 3-loop order \cite{metlitski1}, which are expected to make the self energy non-local, and it remains to be seen if such effects could reduce the maximal Lyapunov exponent. However, since $|p_s|-|p_1|$ is $\mathcal{O}(1)$ at large $N$ (Fig. \ref{fig:LambdaPx}), we expect that the small $\mathcal{O}(1/N)$ corrections to this quantity will not immediately be able to change its sign and thus reduce the maximal Lyapunov exponent.

We also examined a phenomenological model of a critical Fermi surface with dynamic critical exponent $z \leq 3$; quasiparticles re-emerge in such a model for $z < 2$. We found that the maximal chaos was retained for precisely the regime where quasiparticles are absent, $2 < z \leq 3$, although the size of the spactime region for which it occurs shrinks to zero as $z\rightarrow 2$ (Fig. \ref{fig:squeeze}). It is also remarkable that the acceleration of chaos to the maximal rate by the butterfly effect is tied to the destruction of quasiparticles in this model.
When quasiparticles are present, we found that the saddle-point contribution dominated with $\lambda_L \sim T^{2/z} \ll T$, which is parametrically smaller than the maximal rate (Supplementary Information).

{\it Acknolwedgements.} We thank Haoyu Guo and Yingfei Gu for valuable discussions. M.T. and S.S. were supported by the National Science Foundation under Grant No.~DMR-2002850. A.A.P. was supported by the Miller Institute for Basic Research in Science. This work was also supported by the Simons Collaboration on Ultra-Quantum Matter, which is a grant from the Simons Foundation (651440, S.S.).

\bibliography{refs.bib}

\begin{thebibliography}{23}%
\makeatletter
\providecommand \@ifxundefined [1]{%
 \@ifx{#1\undefined}
}%
\providecommand \@ifnum [1]{%
 \ifnum #1\expandafter \@firstoftwo
 \else \expandafter \@secondoftwo
 \fi
}%
\providecommand \@ifx [1]{%
 \ifx #1\expandafter \@firstoftwo
 \else \expandafter \@secondoftwo
 \fi
}%
\providecommand \natexlab [1]{#1}%
\providecommand \enquote  [1]{``#1''}%
\providecommand \bibnamefont  [1]{#1}%
\providecommand \bibfnamefont [1]{#1}%
\providecommand \citenamefont [1]{#1}%
\providecommand \href@noop [0]{\@secondoftwo}%
\providecommand \href [0]{\begingroup \@sanitize@url \@href}%
\providecommand \@href[1]{\@@startlink{#1}\@@href}%
\providecommand \@@href[1]{\endgroup#1\@@endlink}%
\providecommand \@sanitize@url [0]{\catcode `\\12\catcode `\$12\catcode
  `\&12\catcode `\#12\catcode `\^12\catcode `\_12\catcode `\%12\relax}%
\providecommand \@@startlink[1]{}%
\providecommand \@@endlink[0]{}%
\providecommand \url  [0]{\begingroup\@sanitize@url \@url }%
\providecommand \@url [1]{\endgroup\@href {#1}{\urlprefix }}%
\providecommand \urlprefix  [0]{URL }%
\providecommand \Eprint [0]{\href }%
\providecommand \doibase [0]{https://doi.org/}%
\providecommand \selectlanguage [0]{\@gobble}%
\providecommand \bibinfo  [0]{\@secondoftwo}%
\providecommand \bibfield  [0]{\@secondoftwo}%
\providecommand \translation [1]{[#1]}%
\providecommand \BibitemOpen [0]{}%
\providecommand \bibitemStop [0]{}%
\providecommand \bibitemNoStop [0]{.\EOS\space}%
\providecommand \EOS [0]{\spacefactor3000\relax}%
\providecommand \BibitemShut  [1]{\csname bibitem#1\endcsname}%
\let\auto@bib@innerbib\@empty
\bibitem [{\citenamefont {Hartnoll}\ and\ \citenamefont
  {Mackenzie}(2021)}]{Hartnoll:2021ydi}%
  \BibitemOpen
  \bibfield  {author} {\bibinfo {author} {\bibfnamefont {S.~A.}\ \bibnamefont
  {Hartnoll}}\ and\ \bibinfo {author} {\bibfnamefont {A.~P.}\ \bibnamefont
  {Mackenzie}},\ }\bibfield  {title} {\bibinfo {title} {{Planckian Dissipation
  in Metals}},\ }\href@noop {} {\  (\bibinfo {year} {2021})},\ \Eprint
  {https://arxiv.org/abs/2107.07802} {arXiv:2107.07802 [cond-mat.str-el]}
  \BibitemShut {NoStop}%
\bibitem [{\citenamefont {Chowdhury}\ \emph {et~al.}(2021)\citenamefont
  {Chowdhury}, \citenamefont {Georges}, \citenamefont {Parcollet},\ and\
  \citenamefont {Sachdev}}]{Chowdhury:2021qpy}%
  \BibitemOpen
  \bibfield  {author} {\bibinfo {author} {\bibfnamefont {D.}~\bibnamefont
  {Chowdhury}}, \bibinfo {author} {\bibfnamefont {A.}~\bibnamefont {Georges}},
  \bibinfo {author} {\bibfnamefont {O.}~\bibnamefont {Parcollet}},\ and\
  \bibinfo {author} {\bibfnamefont {S.}~\bibnamefont {Sachdev}},\ }\bibfield
  {title} {\bibinfo {title} {{Sachdev-Ye-Kitaev Models and Beyond: A Window
  into Non-Fermi Liquids}},\ }\href@noop {} {\  (\bibinfo {year} {2021})},\
  \Eprint {https://arxiv.org/abs/2109.05037} {arXiv:2109.05037
  [cond-mat.str-el]} \BibitemShut {NoStop}%
\bibitem [{\citenamefont {Sachdev}(1999)}]{qptbook}%
  \BibitemOpen
  \bibfield  {author} {\bibinfo {author} {\bibfnamefont {S.}~\bibnamefont
  {Sachdev}},\ }\href {https://www.cambridge.org/9780521514682} {\emph
  {\bibinfo {title} {{Quantum Phase Transitions}}}}\ (\bibinfo  {publisher}
  {Cambridge University Press},\ \bibinfo {address} {Cambridge, UK},\ \bibinfo
  {year} {1999})\BibitemShut {NoStop}%
\bibitem [{\citenamefont {Bruin}\ \emph {et~al.}(2013)\citenamefont {Bruin},
  \citenamefont {Sakai}, \citenamefont {Perry},\ and\ \citenamefont
  {Mackenzie}}]{bruin}%
  \BibitemOpen
  \bibfield  {author} {\bibinfo {author} {\bibfnamefont {J.~A.~N.}\
  \bibnamefont {Bruin}}, \bibinfo {author} {\bibfnamefont {H.}~\bibnamefont
  {Sakai}}, \bibinfo {author} {\bibfnamefont {R.~S.}\ \bibnamefont {Perry}},\
  and\ \bibinfo {author} {\bibfnamefont {A.~P.}\ \bibnamefont {Mackenzie}},\
  }\bibfield  {title} {\bibinfo {title} {{Similarity of Scattering Rates in
  Metals Showing $T$-Linear Resistivity}},\ }\href
  {https://doi.org/10.1126/science.1227612} {\bibfield  {journal} {\bibinfo
  {journal} {Science}\ }\textbf {\bibinfo {volume} {339}},\ \bibinfo {pages}
  {804} (\bibinfo {year} {2013})}\BibitemShut {NoStop}%
\bibitem [{\citenamefont {Zaanen}(2004)}]{Zaanen}%
  \BibitemOpen
  \bibfield  {author} {\bibinfo {author} {\bibfnamefont {J.}~\bibnamefont
  {Zaanen}},\ }\bibfield  {title} {\bibinfo {title} {Why the temperature is
  high},\ }\href {https://doi.org/10.1038/430512a} {\bibfield  {journal}
  {\bibinfo  {journal} {Nature}\ }\textbf {\bibinfo {volume} {430}},\ \bibinfo
  {pages} {512} (\bibinfo {year} {2004})}\BibitemShut {NoStop}%
\bibitem [{\citenamefont {{Grissonnanche}}\ \emph {et~al.}(2021)\citenamefont
  {{Grissonnanche}}, \citenamefont {{Fang}}, \citenamefont {{Legros}},
  \citenamefont {{Verret}}, \citenamefont {{Lalibert{\'e}}}, \citenamefont
  {{Collignon}}, \citenamefont {{Zhou}}, \citenamefont {{Graf}}, \citenamefont
  {{Goddard}}, \citenamefont {{Taillefer}},\ and\ \citenamefont
  {{Ramshaw}}}]{Gael21}%
  \BibitemOpen
  \bibfield  {author} {\bibinfo {author} {\bibfnamefont {G.}~\bibnamefont
  {{Grissonnanche}}}, \bibinfo {author} {\bibfnamefont {Y.}~\bibnamefont
  {{Fang}}}, \bibinfo {author} {\bibfnamefont {A.}~\bibnamefont {{Legros}}},
  \bibinfo {author} {\bibfnamefont {S.}~\bibnamefont {{Verret}}}, \bibinfo
  {author} {\bibfnamefont {F.}~\bibnamefont {{Lalibert{\'e}}}}, \bibinfo
  {author} {\bibfnamefont {C.}~\bibnamefont {{Collignon}}}, \bibinfo {author}
  {\bibfnamefont {J.}~\bibnamefont {{Zhou}}}, \bibinfo {author} {\bibfnamefont
  {D.}~\bibnamefont {{Graf}}}, \bibinfo {author} {\bibfnamefont {P.~A.}\
  \bibnamefont {{Goddard}}}, \bibinfo {author} {\bibfnamefont {L.}~\bibnamefont
  {{Taillefer}}},\ and\ \bibinfo {author} {\bibfnamefont {B.~J.}\ \bibnamefont
  {{Ramshaw}}},\ }\bibfield  {title} {\bibinfo {title} {{Linear-in temperature
  resistivity from an isotropic Planckian scattering rate}},\ }\href
  {https://doi.org/10.1038/s41586-021-03697-8} {\bibfield  {journal} {\bibinfo
  {journal} {Nature}\ }\textbf {\bibinfo {volume} {595}},\ \bibinfo {pages}
  {667} (\bibinfo {year} {2021})},\ \Eprint {https://arxiv.org/abs/2011.13054}
  {arXiv:2011.13054 [cond-mat.str-el]} \BibitemShut {NoStop}%
\bibitem [{\citenamefont {Maldacena}\ \emph {et~al.}(2016)\citenamefont
  {Maldacena}, \citenamefont {Shenker},\ and\ \citenamefont
  {Stanford}}]{MSS16}%
  \BibitemOpen
  \bibfield  {author} {\bibinfo {author} {\bibfnamefont {J.}~\bibnamefont
  {Maldacena}}, \bibinfo {author} {\bibfnamefont {S.~H.}\ \bibnamefont
  {Shenker}},\ and\ \bibinfo {author} {\bibfnamefont {D.}~\bibnamefont
  {Stanford}},\ }\bibfield  {title} {\bibinfo {title} {{A bound on chaos}},\
  }\href {https://doi.org/10.1007/JHEP08(2016)106} {\bibfield  {journal}
  {\bibinfo  {journal} {JHEP}\ }\textbf {\bibinfo {volume} {08}},\ \bibinfo
  {pages} {106}},\ \Eprint {https://arxiv.org/abs/1503.01409} {arXiv:1503.01409
  [hep-th]} \BibitemShut {NoStop}%
\bibitem [{\citenamefont {Maldacena}\ and\ \citenamefont
  {Stanford}(2016)}]{Maldacena_syk}%
  \BibitemOpen
  \bibfield  {author} {\bibinfo {author} {\bibfnamefont {J.}~\bibnamefont
  {Maldacena}}\ and\ \bibinfo {author} {\bibfnamefont {D.}~\bibnamefont
  {Stanford}},\ }\bibfield  {title} {\bibinfo {title} {{Remarks on the
  Sachdev-Ye-Kitaev model}},\ }\href
  {https://doi.org/10.1103/PhysRevD.94.106002} {\bibfield  {journal} {\bibinfo
  {journal} {Phys. Rev. D}\ }\textbf {\bibinfo {volume} {94}},\ \bibinfo
  {pages} {106002} (\bibinfo {year} {2016})}\BibitemShut {NoStop}%
\bibitem [{\citenamefont {Kitaev}\ and\ \citenamefont
  {Suh}(2018)}]{kitaev2018}%
  \BibitemOpen
  \bibfield  {author} {\bibinfo {author} {\bibfnamefont {A.}~\bibnamefont
  {Kitaev}}\ and\ \bibinfo {author} {\bibfnamefont {S.~J.}\ \bibnamefont
  {Suh}},\ }\bibfield  {title} {\bibinfo {title} {{The soft mode in the
  Sachdev-Ye-Kitaev model and its gravity dual}},\ }\href
  {https://doi.org/10.1007/JHEP05(2018)183} {\bibfield  {journal} {\bibinfo
  {journal} {JHEP}\ }\textbf {\bibinfo {volume} {05}},\ \bibinfo {pages}
  {183}},\ \Eprint {https://arxiv.org/abs/1711.08467} {arXiv:1711.08467
  [hep-th]} \BibitemShut {NoStop}%
\bibitem [{\citenamefont {{Lee}}(2018)}]{SungSik18}%
  \BibitemOpen
  \bibfield  {author} {\bibinfo {author} {\bibfnamefont {S.-S.}\ \bibnamefont
  {{Lee}}},\ }\bibfield  {title} {\bibinfo {title} {{Recent Developments in
  Non-Fermi Liquid Theory}},\ }\href
  {https://doi.org/10.1146/annurev-conmatphys-031016-025531} {\bibfield
  {journal} {\bibinfo  {journal} {Annual Review of Condensed Matter Physics}\
  }\textbf {\bibinfo {volume} {9}},\ \bibinfo {pages} {227} (\bibinfo {year}
  {2018})},\ \Eprint {https://arxiv.org/abs/1703.08172} {arXiv:1703.08172
  [cond-mat.str-el]} \BibitemShut {NoStop}%
\bibitem [{\citenamefont {Patel}\ and\ \citenamefont
  {Sachdev}(2017)}]{patel2017quantum}%
  \BibitemOpen
  \bibfield  {author} {\bibinfo {author} {\bibfnamefont {A.~A.}\ \bibnamefont
  {Patel}}\ and\ \bibinfo {author} {\bibfnamefont {S.}~\bibnamefont
  {Sachdev}},\ }\bibfield  {title} {\bibinfo {title} {{Quantum chaos on a
  critical Fermi surface}},\ }\href {https://doi.org/10.1073/pnas.1618185114}
  {\bibfield  {journal} {\bibinfo  {journal} {Proc. Nat. Acad. Sci.}\ }\textbf
  {\bibinfo {volume} {114}},\ \bibinfo {pages} {1844} (\bibinfo {year}
  {2017})},\ \Eprint {https://arxiv.org/abs/1611.00003} {arXiv:1611.00003
  [cond-mat.str-el]} \BibitemShut {NoStop}%
\bibitem [{\citenamefont {Gu}\ and\ \citenamefont {Kitaev}(2019)}]{gu2019}%
  \BibitemOpen
  \bibfield  {author} {\bibinfo {author} {\bibfnamefont {Y.}~\bibnamefont
  {Gu}}\ and\ \bibinfo {author} {\bibfnamefont {A.}~\bibnamefont {Kitaev}},\
  }\bibfield  {title} {\bibinfo {title} {{On the relation between the magnitude
  and exponent of OTOCs}},\ }\href {https://doi.org/10.1007/JHEP02(2019)075}
  {\bibfield  {journal} {\bibinfo  {journal} {JHEP}\ }\textbf {\bibinfo
  {volume} {02}},\ \bibinfo {pages} {075}},\ \Eprint
  {https://arxiv.org/abs/1812.00120} {arXiv:1812.00120 [hep-th]} \BibitemShut
  {NoStop}%
\bibitem [{\citenamefont {Esterlis}\ \emph {et~al.}(2021)\citenamefont
  {Esterlis}, \citenamefont {Guo}, \citenamefont {Patel},\ and\ \citenamefont
  {Sachdev}}]{esterlis2021}%
  \BibitemOpen
  \bibfield  {author} {\bibinfo {author} {\bibfnamefont {I.}~\bibnamefont
  {Esterlis}}, \bibinfo {author} {\bibfnamefont {H.}~\bibnamefont {Guo}},
  \bibinfo {author} {\bibfnamefont {A.~A.}\ \bibnamefont {Patel}},\ and\
  \bibinfo {author} {\bibfnamefont {S.}~\bibnamefont {Sachdev}},\ }\bibfield
  {title} {\bibinfo {title} {{Large $N$ theory of critical Fermi surfaces}},\
  }\href {https://doi.org/10.1103/PhysRevB.103.235129} {\bibfield  {journal}
  {\bibinfo  {journal} {Phys. Rev. B}\ }\textbf {\bibinfo {volume} {103}},\
  \bibinfo {pages} {235129} (\bibinfo {year} {2021})},\ \Eprint
  {https://arxiv.org/abs/2103.08615} {arXiv:2103.08615 [cond-mat.str-el]}
  \BibitemShut {NoStop}%
\bibitem [{\citenamefont {Aldape}\ \emph {et~al.}(2020)\citenamefont {Aldape},
  \citenamefont {Cookmeyer}, \citenamefont {Patel},\ and\ \citenamefont
  {Altman}}]{Aldape2020}%
  \BibitemOpen
  \bibfield  {author} {\bibinfo {author} {\bibfnamefont {E.~E.}\ \bibnamefont
  {Aldape}}, \bibinfo {author} {\bibfnamefont {T.}~\bibnamefont {Cookmeyer}},
  \bibinfo {author} {\bibfnamefont {A.~A.}\ \bibnamefont {Patel}},\ and\
  \bibinfo {author} {\bibfnamefont {E.}~\bibnamefont {Altman}},\ }\bibfield
  {title} {\bibinfo {title} {Solvable theory of a strange metal at the
  breakdown of a heavy {F}ermi liquid},\ }\href
  {https://arxiv.org/abs/2012.00763} {\  (\bibinfo {year} {2020})},\ \Eprint
  {https://arxiv.org/abs/2012.00763} {arXiv:2012.00763 [cond-mat.str-el]}
  \BibitemShut {NoStop}%
\bibitem [{\citenamefont {Lee}(2009)}]{Lee:2009epi}%
  \BibitemOpen
  \bibfield  {author} {\bibinfo {author} {\bibfnamefont {S.-S.}\ \bibnamefont
  {Lee}},\ }\bibfield  {title} {\bibinfo {title} {{Low energy effective theory
  of Fermi surface coupled with U(1) gauge field in 2+1 dimensions}},\ }\href
  {https://doi.org/10.1103/PhysRevB.80.165102} {\bibfield  {journal} {\bibinfo
  {journal} {Phys. Rev. B}\ }\textbf {\bibinfo {volume} {80}},\ \bibinfo
  {pages} {165102} (\bibinfo {year} {2009})},\ \Eprint
  {https://arxiv.org/abs/0905.4532} {arXiv:0905.4532 [cond-mat.str-el]}
  \BibitemShut {NoStop}%
\bibitem [{\citenamefont {{Metlitski}}\ and\ \citenamefont
  {{Sachdev}}(2010)}]{metlitski1}%
  \BibitemOpen
  \bibfield  {author} {\bibinfo {author} {\bibfnamefont {M.~A.}\ \bibnamefont
  {{Metlitski}}}\ and\ \bibinfo {author} {\bibfnamefont {S.}~\bibnamefont
  {{Sachdev}}},\ }\bibfield  {title} {\bibinfo {title} {{Quantum phase
  transitions of metals in two spatial dimensions. I. Ising-nematic order}},\
  }\href {https://doi.org/10.1103/PhysRevB.82.075127} {\bibfield  {journal}
  {\bibinfo  {journal} {Phys. Rev. B}\ }\textbf {\bibinfo {volume} {82}},\
  \bibinfo {eid} {075127} (\bibinfo {year} {2010})},\ \Eprint
  {https://arxiv.org/abs/1001.1153} {arXiv:1001.1153 [cond-mat.str-el]}
  \BibitemShut {NoStop}%
\bibitem [{Note1()}]{Note1}%
  \BibitemOpen
  \bibinfo {note} {For the single patch theory, the values of $c_f,c_b$ are
  given by $c_f=(M/N)2^{5/3}g^{4/3}/3^{3/2}$ and $c_b = g^2/(8\pi
  )$.}\BibitemShut {Stop}%
\bibitem [{Note2()}]{Note2}%
  \BibitemOpen
  \bibinfo {note} {The numerical values of $|p_s|$ and $|p_1|$ are provided for
  chosen parameters $g=0.5$, $\Lambda = 50$, $T=1$, and $z=3/2$. See
  Supplementary Information for more details.}\BibitemShut {Stop}%
\bibitem [{\citenamefont {Gu}\ \emph {et~al.}(2017)\citenamefont {Gu},
  \citenamefont {Qi},\ and\ \citenamefont {Stanford}}]{Gu2017XLQ}%
  \BibitemOpen
  \bibfield  {author} {\bibinfo {author} {\bibfnamefont {Y.}~\bibnamefont
  {Gu}}, \bibinfo {author} {\bibfnamefont {X.-L.}\ \bibnamefont {Qi}},\ and\
  \bibinfo {author} {\bibfnamefont {D.}~\bibnamefont {Stanford}},\ }\bibfield
  {title} {\bibinfo {title} {Local criticality, diffusion and chaos in
  generalized sachdev-ye-kitaev models},\ }\href
  {https://doi.org/10.1007/JHEP05(2017)125} {\bibfield  {journal} {\bibinfo
  {journal} {Journal of High Energy Physics}\ }\textbf {\bibinfo {volume}
  {2017}},\ \bibinfo {pages} {125} (\bibinfo {year} {2017})}\BibitemShut
  {NoStop}%
\bibitem [{\citenamefont {Kim}\ \emph {et~al.}(2021)\citenamefont {Kim},
  \citenamefont {Altman},\ and\ \citenamefont {Cao}}]{Altman2021}%
  \BibitemOpen
  \bibfield  {author} {\bibinfo {author} {\bibfnamefont {J.}~\bibnamefont
  {Kim}}, \bibinfo {author} {\bibfnamefont {E.}~\bibnamefont {Altman}},\ and\
  \bibinfo {author} {\bibfnamefont {X.}~\bibnamefont {Cao}},\ }\bibfield
  {title} {\bibinfo {title} {Dirac fast scramblers},\ }\href
  {https://doi.org/10.1103/PhysRevB.103.L081113} {\bibfield  {journal}
  {\bibinfo  {journal} {Phys. Rev. B}\ }\textbf {\bibinfo {volume} {103}},\
  \bibinfo {pages} {L081113} (\bibinfo {year} {2021})}\BibitemShut {NoStop}%
\bibitem [{\citenamefont {Turiaci}\ and\ \citenamefont
  {Verlinde}(2016)}]{turiaci2016cft}%
  \BibitemOpen
  \bibfield  {author} {\bibinfo {author} {\bibfnamefont {G.~J.}\ \bibnamefont
  {Turiaci}}\ and\ \bibinfo {author} {\bibfnamefont {H.}~\bibnamefont
  {Verlinde}},\ }\bibfield  {title} {\bibinfo {title} {On {C}{F}{T} and quantum
  chaos},\ }\href {https://doi.org/10.1007/JHEP12(2016)110} {\bibfield
  {journal} {\bibinfo  {journal} {Journal of High Energy Physics}\ }\textbf
  {\bibinfo {volume} {2016}},\ \bibinfo {pages} {1} (\bibinfo {year}
  {2016})}\BibitemShut {NoStop}%
\bibitem [{\citenamefont {Haehl}\ and\ \citenamefont
  {Rozali}(2018)}]{haehl2018effective}%
  \BibitemOpen
  \bibfield  {author} {\bibinfo {author} {\bibfnamefont {F.~M.}\ \bibnamefont
  {Haehl}}\ and\ \bibinfo {author} {\bibfnamefont {M.}~\bibnamefont {Rozali}},\
  }\bibfield  {title} {\bibinfo {title} {Effective field theory for chaotic
  {C}{F}{T}s},\ }\href {https://doi.org/10.1007/JHEP10(2018)118} {\bibfield
  {journal} {\bibinfo  {journal} {Journal of High Energy Physics}\ }\textbf
  {\bibinfo {volume} {2018}},\ \bibinfo {pages} {1} (\bibinfo {year}
  {2018})}\BibitemShut {NoStop}%
\bibitem [{\citenamefont {Haehl}\ \emph {et~al.}(2019)\citenamefont {Haehl},
  \citenamefont {Reeves},\ and\ \citenamefont
  {Rozali}}]{haehl2019reparametrization}%
  \BibitemOpen
  \bibfield  {author} {\bibinfo {author} {\bibfnamefont {F.~M.}\ \bibnamefont
  {Haehl}}, \bibinfo {author} {\bibfnamefont {W.}~\bibnamefont {Reeves}},\ and\
  \bibinfo {author} {\bibfnamefont {M.}~\bibnamefont {Rozali}},\ }\bibfield
  {title} {\bibinfo {title} {Reparametrization modes, shadow operators, and
  quantum chaos in higher-dimensional {C}{F}{T}s},\ }\href
  {https://doi.org/10.1007/JHEP11(2019)102} {\bibfield  {journal} {\bibinfo
  {journal} {Journal of High Energy Physics}\ }\textbf {\bibinfo {volume}
  {2019}},\ \bibinfo {pages} {1} (\bibinfo {year} {2019})}\BibitemShut
  {NoStop}%
\end{thebibliography}%

\newpage


\section*{Supplementary material}

# Supplementary Information.
# Maximal quantum chaos of the critical Fermi surface

Maria Tikhanovskaya,[1] Subir Sachdev,[1, 2] and Aavishkar A. Patel[3]

[1]*Department of Physics, Harvard University, Cambridge MA 02138, USA*
[2]*School of Natural Sciences, Institute for Advanced Study, Princeton, NJ-08540, USA*
[3]*Department of Physics, University of California Berkeley, Berkeley CA 94720, USA*

## I. DETAILS OF THE EIGENVALUE EQUATION

In what follows we derive the one dimensional eigenvalue equation that is solved in the main text. A similar analysis of the Feynman rules for the OTOC was done before *e.g.* in [1–3]. We will present explicit derivations for each diagram and describe how the final structure is obtained. The analysis is first done for the single patch theory, and we later on generalize to two antipodal patches, and show that our results and conclusions do not change.

The object of interest is the regulated squared anticommutator of the fermion operators

$$\mathcal{C}_{\bm{x}}(t,0) = \frac{1}{N^2}\theta(t)\sum_{n,m=1}^{N}\text{Tr}\left[e^{-\beta H/2}\{\psi_n(\mathbf{x},t),\psi_m^\dagger(0)\}e^{-\beta H/2}\{\psi_n(\mathbf{x},t),\psi_m^\dagger(0)\}^\dagger\right], \quad (1)$$

where $\psi(\mathbf{x},t) = U_I^\dagger \psi_0(\mathbf{x},t) U_I$ and $U_I$ is the interaction picture time evolution operator. The notation $\psi(0) \equiv \psi(\mathbf{0},0)$ is introduced for simplicity. In the following, we drop the subscript "0" on $\psi$, assuming that the fields are free.

A Taylor expansion of the evolution operator up to second order reads

$$U_I = \exp\left(\frac{i}{N}\sum_{ijl}\int_\mathbf{y}\int_0^t ds\, g_{ijl}\psi_i^\dagger(\mathbf{y},s)\psi_j(\mathbf{y},s)\phi_l(\mathbf{y},s)\right) = 1 + \frac{i}{N}\sum_{ijl}\int_\mathbf{y}\int_0^t ds\, g_{ijl}\psi_i^\dagger(\mathbf{y},s)\psi_j(\mathbf{y},s)\phi_l(\mathbf{y},s) \quad (2)$$

$$+ \left(\frac{i}{N}\right)^2\sum_{\text{all indecies}}\int_{\mathbf{y},\mathbf{y}'}\int_0^t ds\int_0^s ds'\, g_{ijl}g_{i'j'l'}\phi_l(\mathbf{y},s)\phi_{l'}(\mathbf{y}',s')\psi_i^\dagger(\mathbf{y},s)\psi_j(\mathbf{y},s)\psi_{i'}^\dagger(\mathbf{y}',s')\psi_{j'}(\mathbf{y}',s') + \ldots.$$

An expansion of its conjugate is therefore given by

$$U_I^\dagger = \exp\left(-\frac{i}{N}\sum_{ijl}\int_\mathbf{y}\int_0^t ds\, g_{ijl}\psi_i^\dagger(\mathbf{y},s)\psi_j(\mathbf{y},s)\phi_l(\mathbf{y},s)\right) = 1 - \frac{i}{N}\sum_{ijl}\int_\mathbf{y}\int_0^t ds\, g_{ijl}\psi_i^\dagger(\mathbf{y},s)\psi_j(\mathbf{y},s)\phi_l(\mathbf{y},s) \quad (3)$$

$$+ \left(\frac{-i}{N}\right)^2\sum_{\text{all indecies}}\int_{\mathbf{y},\mathbf{y}'}\int_0^t ds\int_0^s ds'\, g_{ijl}g_{i'j'l'}\phi_l(\mathbf{y},s)\phi_{l'}(\mathbf{y}',s')\psi_i^\dagger(\mathbf{y},s)\psi_j(\mathbf{y},s)\psi_{i'}^\dagger(\mathbf{y}',s')\psi_{j'}(\mathbf{y}',s') + \ldots.$$

We use the following definitions of retarded Green's functions and the symmetrized Wightman functions for fermions and bosons:

$$G^R(\mathbf{x},t)\delta_{a,b} = -i\theta(t)\langle\{\psi_a(\mathbf{x},t),\psi_b^\dagger(0)\}\rangle, \quad (4)$$

$$D^R(\mathbf{x},t)\delta_{a,b} = -i\theta(t)\langle[\phi_a(\mathbf{x},t),\phi_b(0)]\rangle, \quad (5)$$

$$G^W(\mathbf{x},t)\delta_{a,b} = \text{Tr}[\rho^{1/2}\psi_a(\mathbf{x},t)\rho^{1/2}\psi_b^\dagger(0)], \quad (6)$$

$$D^W(\mathbf{x},t)\delta_{a,b} = \text{Tr}[\rho^{1/2}\phi_a(\mathbf{x},t)\rho^{1/2}\phi_b(0)]. \quad (7)$$

where $\rho = e^{-\beta H}$.

**Leading contribution.**

The zeroth order Taylor expansion of the evolution operator (2), along with the above definition of the retarded Green's function for fermions (4), gives us the leading contribution to the expression for the squared anticommutator

$$\mathcal{C}_{\bm{x}}^{(0)}(t,0) = \frac{1}{N^2}\theta(t)\sum_{n,m=1}^{N}\text{Tr}\left[\rho^{1/2}\{\psi_m(\mathbf{x},t),\psi_n^\dagger(0)\}\rho^{1/2}\{\psi_m(\mathbf{x},t),\psi_n^\dagger(0)\}^\dagger\right]$$

$$= \frac{1}{N}(iG^R(\mathbf{x},t))(-iG^{R*}(\mathbf{x},t)) = \frac{1}{N}|G^R(\mathbf{x},t)|^2,$$

Diagrammatically, we represent it simply as

$$\mathcal{C}_{\boldsymbol{x}}^{(0)}(t,0) = \quad. \tag{8}$$

**First order.**

To derive the eigenvalue equation at the first order, we note that only the product of even numbers of the coupling constants $g_{ijl}$ gives a non-zero result. The combination of the first term on one of the time folds, and the expansion (2)-(3) to higher orders on the other time fold leads to corrections self energy corrections to the Green's function. Thus we are left with the expansion (2)-(3) to the first order on both time folds, and we obtain

$$\mathcal{C}_{\boldsymbol{x}}^{(1)}(t,0) = \frac{1}{N^4}\theta(t) \int_{s,s'} \int_{\mathbf{y},\mathbf{y}'} \sum_{\text{all indecies}} \text{Tr}[\rho^{1/2} g_{ijl}\phi_l(\mathbf{y},s)\{[\psi_m(\mathbf{x},t),\psi_i^\dagger(\mathbf{y},s)\psi_j(\mathbf{y},s)],\psi_n^\dagger(0)\}$$
$$\times \rho^{1/2} g_{i'j'l'} \phi_{l'}(\mathbf{y}',s')\{[\psi_m(\mathbf{x},t),\psi_{i'}^\dagger(\mathbf{y}',s')\psi_{j'}(\mathbf{y}',s')],\psi_n^\dagger(0)\}^\dagger]$$

Using definitions (4)-(7), and counting the powers of $N$ for each term, we obtain the following expression

$$\mathcal{C}_{\boldsymbol{x}}^{(1)}(t,0) = \frac{g^2}{N} \int_{s,s'} \int_{\mathbf{y},\mathbf{y}'} G^R(\mathbf{x}-\mathbf{y},t-s)G^R(\mathbf{y},s)D^W(\mathbf{y}-\mathbf{y}',s-s')G^{R*}(\mathbf{x}-\mathbf{y}',t-s')G^{R*}(\mathbf{y}',s'),$$

which gives the "rung" diagram

$$\mathcal{C}_{\boldsymbol{x}}^{(1)}(t,0) = \quad. \tag{9}$$

**Second order.**

At the second order we find three terms, but only one of them is both nonzero at large $N$ and is not two ladders of type (9). This term is a "box" diagram and is similar to the one found in [2] and reads

$$\mathcal{C}_{\boldsymbol{x}}^{(2)}(t,0) = \frac{g^4}{N}\int_{\{\mathbf{y}\}}\int_0^t ds_1 \int_0^{s_1} ds_2 \int_0^t ds_3 \int_0^{s_3} ds_4 G^R(\mathbf{x}-\mathbf{y}_1,t-s_1)D^R(\mathbf{y}_1-\mathbf{y}_2,s_1-s_2)G^W(\mathbf{y}_1-\mathbf{y}_3,s_1-s_3)$$
$$\times G^W(\mathbf{y}_4-\mathbf{y}_2,s_4-s_2)G^R(\mathbf{y}_2,s_2)G^{R*}(\mathbf{x}-\mathbf{y}_3,t-s_3)D^{R*}(\mathbf{y}_3-\mathbf{y}_4,s_3-s_4)G^{R*}(\mathbf{y}_4,s_4)$$

The diagram is

$$\mathcal{C}_{\boldsymbol{x}}^{(2)}(t,0) = \quad. \tag{10}$$

In the large $N$ limit, we don't find any new types of diagrams other than boxes and rungs contributing to (1), after expanding the unitary operator (2) to higher orders.

**The eigenvalue equation.**

We now evaluate the ladder sum and work with the following function:

$$\mathcal{C}(\omega) = \frac{1}{N}\int \frac{d^3k}{(2\pi)^3}\mathcal{C}(k,\omega), \tag{11}$$

where $\mathcal{C}(k,\omega)$ is the Fourier transform of $\mathcal{C}_{\boldsymbol{x}}(t,0)$ with respect to $\boldsymbol{x}$ and $t$.

The Bethe-Saltpeter equation (Fig. 3 of the main text) then reads

$$\mathcal{C}(k,\omega) = G^R(k)G^{R*}(k-\omega)\left[1 + \int \frac{d^3k'}{(2\pi)^3}(g^2 D^W(k-k') + K_2(k,k',\omega))\mathcal{C}(k',\omega)\right]. \tag{12}$$

Each of the terms corresponds to Fourier transform of the perturbative expansion found above. The exponential growth of the squared anticommutator in the chaos regime is expected if the eigenvalue equation is invariant under adding a ladder, and we can then write

$$\mathcal{C}(k,\omega) = G^R(k)G^{R*}(k-\omega)\int \frac{d^3k'}{(2\pi)^3}(g^2 D^W(k-k') + K_2(k,k',\omega))\mathcal{C}(k',\omega), \tag{13}$$

where the function $K_2(k, k', \omega)$ is the kernel of the "box" diagram found above (10), which in Fourier space reads

$$K_2(k, k', \omega) = g^4 \int \frac{d^3 k_1}{(2\pi)^3} D^R(k_1) D^{R*}(k_1 - \omega) G^W(k - k_1) G^W(k' - k_1). \tag{14}$$

We can now explicitly compute each term in (13). The fermion and boson Green's functions the single patch theory [4] are given by

$$G^R(\mathbf{k}, \omega) = \frac{1}{k_x + k_y^2 - \Sigma^R(\omega)}, \tag{15}$$

$$G^W(\mathbf{k}, \omega) = -\frac{\mathrm{Im}\Sigma^R(\omega)}{\cosh\frac{\beta\omega}{2}\left([(k_x + k_y^2) + \mathrm{Re}\Sigma^R(\omega)]^2 + [\mathrm{Im}\Sigma^R(\omega)]^2\right)}, \tag{16}$$

$$D^R(\mathbf{q}, \Omega \neq 0) = \frac{|q_y|}{|q_y|^3 + m^2 - ic_b\Omega}, \tag{17}$$

$$D^W(\mathbf{q}, \Omega) = \frac{c_b \Omega |q_y|}{\sinh\frac{\beta\Omega}{2}((|q_y|^3 + m^2)^2 + c_b^2 \Omega^2)}, \tag{18}$$

where we added a small mass term in the boson Green's function as an IR regulator that will eventually be carefully taken to zero, and

$$\Sigma^R(\omega) = ic_f T^{2/3} H_{1/3}\left(\frac{-i\omega - \pi T}{2\pi T}\right) \tag{19}$$

The function $H_{1/3}(z) = \zeta(1/3) - \zeta(1/3, z+1)$ is the Harmonic number function, as noted in the main text, and the constants take the values $c_f = 2^{5/3} g^{4/3}/(3\sqrt{3})$ and $c_b = g^2/(8\pi)$ in the single patch theory.

Since our theory has sliding symmetry about the patch Fermi surface, the function $\mathcal{C}(k, \omega)$ depends on momentum as $\mathcal{C}(k, \omega) = \mathcal{C}(k_x + k_y^2, k_0, \omega)$. Our main goal further is to express the equation in terms of a momentum independent eigenfunction [2]

$$\tilde{\mathcal{C}}(k_0', \omega) = \int \frac{dk_x'}{2\pi} \mathcal{C}(k_x', k_0', \omega), \tag{20}$$

which we will see is possible to do because of the special momentum dependence of $\mathcal{C}$ induced by the sliding symmetry.

We now compute each term in (13). The diagonal term in the equation is simply

$$\int \frac{dk_x}{2\pi} G^R(k) G^{R*}(k - \omega) = \frac{i}{\Sigma^R(k_0) - \Sigma^{R*}(k_0 - \omega)}$$
$$= \frac{1}{c_f T^{2/3}} \frac{1}{H_{1/3}\left(\frac{-ik_0 - \pi T}{2\pi T}\right) + H_{1/3}\left(\frac{-i(\omega - k_0) - \pi T}{2\pi T}\right)}. \tag{21}$$

The "first order" term can be written as

$$g^2 \int \frac{d^3 k'}{(2\pi)^3} D^W(k - k') \mathcal{C}(k', \omega) \tag{22}$$
$$= g^2 c_b \int \frac{dk_0' dk_y'}{(2\pi)^2} \frac{(k_0 - k_0')|k_y'|}{(k_y'^3 + m^2)^2 + c_b^2(k_0 - k_0')^2} \frac{\tilde{\mathcal{C}}(k_0', \omega)}{\sinh\frac{k_0 - k_0'}{2T}}.$$

The second term in the eigenvalue equation, i.e. the $K_2$ - term can be evaluated as done in Ref. [2]

$$\int \frac{d^3 k'}{(2\pi)^3} K_2(k, k', \omega) \mathcal{C}(k', \omega) = g^4 \int \frac{d^3 k_1 d^3 k'}{(2\pi)^6} D^R(k_1) D^{R*}(k_1 - \omega) G^W(k - k_1) G^W(k' - k_1) \mathcal{C}(k', \omega)$$
$$= \frac{g^{4/3} 4\pi^{4/3}}{3\sqrt{3}} \int \frac{dk_0' dk_{01}}{(2\pi)^2} \frac{(ik_{01} + (-ik_{01})^{2/3}(i(k_{01} - \omega)))}{k_{01}(i(k_{01} - \omega))^{1/3}(2k_{01} - \omega)} \frac{\tilde{\mathcal{C}}(k_0', \omega)}{\cosh\frac{k_0 - k_{01}}{2T} \cosh\frac{k_0' - k_{01}}{2T}}. \tag{23}$$

The integration over frequencies has to be done numerically. Combining the above equations, we obtain the eigenvalue equation that we solve in the main text:

$$\left[c_f T^{2/3}\left(H_{1/3}\left(\frac{-ik_0 - \pi T}{2\pi T}\right) + H_{1/3}\left(\frac{-i(\omega - k_0) - \pi T}{2\pi T}\right)\right) + 2\mu(T)\right]\tilde{\mathcal{C}}(k_0,\omega) \quad (24)$$

$$= g^2 \int \frac{dk'_0 dk'_y}{(2\pi)^2} \frac{c_b(k_0 - k'_0)|k'_y|}{(|k'_y|^3 + m^2)^2 + c_b^2(k_0 - k'_0)^2} \frac{\tilde{\mathcal{C}}(k'_0,\omega)}{\sinh\frac{k_0 - k'_0}{2T}}$$

$$+ \frac{g^{4/3} 4\pi^{4/3}}{3\sqrt{3}} \int \frac{dk'_0 dk_{01}}{(2\pi)^2} \frac{(ik_{01} + (-ik_{01})^{2/3}(i(k_{01}-\omega)))}{k_{01}(i(k_{01}-\omega))^{1/3}(2k_{01}-\omega)} \frac{\tilde{\mathcal{C}}(k'_0,\omega)}{\cosh\frac{k_0-k_{01}}{2T}\cosh\frac{k'_0-k_{01}}{2T}}.$$

We then numerically solve the equation following Appendix D in [2].

Our numerical procedure is as follows. We first perform the integration over $k_y$ with a small finite mass term $m = 0.02$. We then discretize the integration over $k'_0$ as well as the variable $k_0$ in the region $k_0, k'_0 \in [-15, 15]$ with the step size $dk_0 = 0.005$. We solve the eigenvalue equation for different $\omega$ on the positive imaginary axis, and look for $\omega$ at which eigenvalue of the equation (24) is closest to zero. Whenever we obtain a solution with multiple eigenvalues, we choose the largest one since it corresponds to the Lyapunov exponent $\lambda_L$. As a check, we find the value $\lambda_L \approx 2.48T$ at $p_x = 0$, which is exactly the result that was found in [2].

**Antipodal patches.**

We now consider the evolution of the OTOC in the non-chiral theory with two antipodal patches. We will show that our conclusions about maximal quantum chaos are unaffected in the large $N$ limit by the interactions induced between the two patches.

The action for the two patch model ((1) of the main text) tells us that the fermions in the $\pm$ patches disperse as $\pm k_x + k_y^2$ respectively. We therefore expect that the eigenfunctions for the two patches $\pm$ are given by $\mathcal{C}_\pm(k, k_0, \omega) = \mathcal{C}(\pm k_x + k_y^2, k_0, \omega)$, which is obvious at the non-interacting level (i.e. (8)), and can easily be shown to be self-consistent when interactions are included in the Bethe-Salpeter equation.

In addition to the $K_2$ term in the eigenvalue equation, we now have a term that couples fermions from opposite patches at the two ends of the "box" diagrams. For the $+$ patch, it is given by (25),

$$\mathcal{C}'^{(2)}_x(t,0) = \quad\text{(diagram)}, \quad (25)$$

which yields

$$\int \frac{d^3 k'}{(2\pi)^3} K'_2(k, k', \omega)\mathcal{C}_-(k', \omega) = \int \frac{d^3 k'}{(2\pi)^3} K'_2(k, k'_0, k'_x, k'_y, \omega)\mathcal{C}(-k'_x + k'^2_y, k_0, \omega). \quad (26)$$

Then, by exploiting the sliding symmetry and shifting $k'_x \to k'_x + k'^2_y$, we can complete the internal momentum $(k_{x1}, k_{y1})$ integrals just like Ref. [2] did with the $K_2$ term, followed by the $(k'_x, k'_y)$ integrals. This yields

$$\int \frac{d^3 k'}{(2\pi)^3} K'_2(k, k', \omega)\mathcal{C}_-(k', \omega) = g^4 \int \frac{d^3 k_1 d^3 k'}{(2\pi)^6} D^R(k_1) D^{R*}(k_1 - \omega) G^W_+(k-k_1) G^W_-(k'-k_1)\mathcal{C}_-(k',\omega)$$

$$= \frac{g^{4/3} 4\pi^{4/3}}{2^{4/3} 3\sqrt{3}} \int \frac{dk'_0 dk_{01}}{(2\pi)^2} \frac{(ik_{01} + (-ik_{01})^{2/3}(i(k_{01}-\omega)))}{k_{01}(i(k_{01}-\omega))^{1/3}(2k_{01}-\omega)} \frac{\tilde{\mathcal{C}}(k'_0,\omega)}{\cosh\frac{k_0-k_{01}}{2T}\cosh\frac{k'_0-k_{01}}{2T}}, \quad (27)$$

where $G^W_\pm$ are the Wightman fermion Green's functions for the $\pm$ patches respectively. This is just proportional to the action of the $K_2$ term on $\mathcal{C}_+$ (23). The sum of the action of the $K_2$ and $K'_2$ terms therefore yields the RHS of (23) divided by a factor of $2^{1/3}$. However, since the two patch value of $c_b$ is twice its one patch value, and the two patch value of $c_f$ is $1/2^{1/3}$ times its one patch value, this extra factor of $1/2^{1/3}$ cancels out in the equivalent of (24) as $m$ is taken to 0, leaving us with the same final one dimensional integral equation to be solved numerically.

Therefore, at $ip_x = 0$, there is no difference between the solutions for the one-patch and two-patch OTOCs. For non-zero external momentum, we can see that if we apply $+ip_x$ to the $+$ patch and $-ip_x$ to the $-$ patch, corresponding to a pair of chiral scrambling modes traveling with the same speed but in opposite directions (one coming from each patch), we get the same $ip_x$ dependence for the OTOC as well. The only difference is that the value of $ip_x$ is rescaled by a factor of $2^{1/3}$, and therefore the butterfly velocity is rescaled by a factor of $1/2^{1/3}$. However, since this merely corresponds to a rescaling of the $x$-axes of Figs. 4, 5 of the main text, our conclusions regarding maximal chaos remain unchanged.

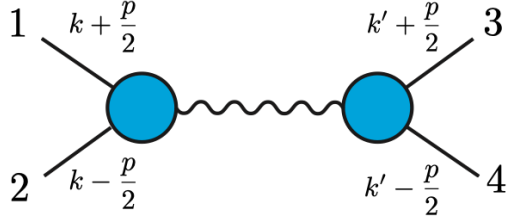

FIG. 1. Diagram representing the ansatz (30). The wavy line represents the scramblon. The figure is adapted from [6] for the momentum-dependent case.

## II.  GENERALIZATION OF THE LADDER IDENTITY

In this section we discuss the generalization of the single mode ansatz [5–7] to the spatially dependent case, and show that the result obtained by Gu and Kitaev [6] holds.

The "regular" OTOC we are considering consists of connected and disconnected parts

$$\text{OTOC}_{\mathbf{x}}(t_1,t_2,t_3,t_4) = \frac{1}{N^2} \sum_{n,m=1}^{N} (\langle \rho^{1/4} \psi_n(\mathbf{x},t_1) \rho^{1/4} \psi_m^\dagger(\mathbf{0},t_3) \rho^{1/4} \psi_n^\dagger(\mathbf{x},t_2) \rho^{1/4} \psi_m(\mathbf{0},t_4) \rangle$$
$$+ \langle \psi_n(\mathbf{x},t_1) \psi_n^\dagger(\mathbf{x},t_2) \rangle \langle \psi_m^\dagger(\mathbf{0},t_3) \psi_m(\mathbf{0},t_4) \rangle). \qquad (28)$$

We consider the Fourier transform of the connected part:

$$\text{OTOC}_{\mathbf{x}}(t_1,t_2,t_3,t_4) = \int_{\mathbf{p},\mathbf{k},\mathbf{k}'} e^{i\mathbf{p}\mathbf{x}} \text{OTOC}_{\mathbf{p}},(t_1,t_2,t_3,t_4;\mathbf{k},\mathbf{k}') \qquad (29)$$

and use the assumption that the single mode ansatz works for every momentum eigenmode. The OTOC then has the following form

$$\text{OTOC}_{\mathbf{p}}(t_1,t_2,t_3,t_4;\mathbf{k},\mathbf{k}') \approx \frac{e^{\kappa(\mathbf{p})(t_1+t_2-t_3-t_4)/2}}{C(\mathbf{p})} \Upsilon^R_{\mathbf{p}}(t_{12},\mathbf{k}) \Upsilon^A_p(t_{34},\mathbf{k}'), \qquad (30)$$

where we assume that the times are well separated $s = (t_1+t_2-t_3-t_4)/2 \gg \kappa(\mathbf{p})^{-1}$, the exponent represents the "scramblon", the momentum-dependent function $C(\mathbf{p})$ is to be determined below from the consistency condition, and $\Upsilon^{A,R}_p(t_{ij},\mathbf{k})$ are the momentum-dependent vertex functions. The diagram that represents the ansatz is shown in Fig. 1.

Let us consider the following form of the regular OTOC, where we distinguish 2 ladders somewhere in the middle. We can then write down the self consistency condition as follows

$$\text{OTOC}_{\mathbf{p}}(t_1,t_2,t_3,t_4;\mathbf{k},\mathbf{k}') \approx \int_{t_5,t_6,t_7,t_8;\mathbf{q},\mathbf{q}'} \text{OTOC}^R_{\mathbf{p}}(t_1,t_2,t_5,t_6;\mathbf{k},\mathbf{q}) \times \boxed{\phantom{XXX}} \times \text{OTOC}_{\mathbf{p}}(t_7,t_8,t_3,t_4;\mathbf{q}',\mathbf{k}').$$

BOX

The points $t_5$ and $t_6$ have the same corresponding imaginary time components as $t_1$ and $t_2$. We therefore need to shift the coordinates on the Keldysh contour by $\pm i\pi/2$ to either direction as shown in Fig.2. There are two choices

$$\text{OTOC}_{\mathbf{p},A} = \text{OTOC}_{\mathbf{p}}\left(t_1,t_2,t_5-i\frac{\pi}{2},t_6-i\frac{\pi}{2};\mathbf{k},\mathbf{k}'\right) \approx \frac{e^{\kappa(\mathbf{p})(t_1+t_2-t_5-t_6+i\pi)/2}}{C(\mathbf{p})} \Upsilon^R_{\mathbf{p}}(t_{12},\mathbf{k}) \Upsilon^A_{\mathbf{p}}(t_{56},\mathbf{k}'),$$

$$\text{OTOC}_{\mathbf{p},A'} = -\text{OTOC}_{\mathbf{p}}\left(t_1,t_2,t_6+i\frac{\pi}{2},t_5+i\frac{\pi}{2};\mathbf{k},\mathbf{k}'\right) \approx \frac{e^{\kappa(\mathbf{p})(t_1+t_2-t_5-t_6-i\pi)/2}}{C(\mathbf{p})} \Upsilon^R_{\mathbf{p}}(t_{12},\mathbf{k}) \Upsilon^A_{\mathbf{p}}(t_{56},\mathbf{k}'),$$

that we need to add together. The other OTOC in (31) is regular because the points $t_3$, $t_4$ and $t_7$, $t_8$ have different imaginary parts, and we obtain

$$\text{OTOC}_{\mathbf{p}}(t_7,t_8,t_3,t_4;\mathbf{k},\mathbf{k}') \approx \frac{e^{\kappa(\mathbf{p})(t_7+t_8-t_3-t_4)/2}}{C(\mathbf{p})} \Upsilon^R_{\mathbf{p}}(t_{78},\mathbf{k}) \Upsilon^A_{\mathbf{p}}(t_{34},\mathbf{k}'). \qquad (31)$$

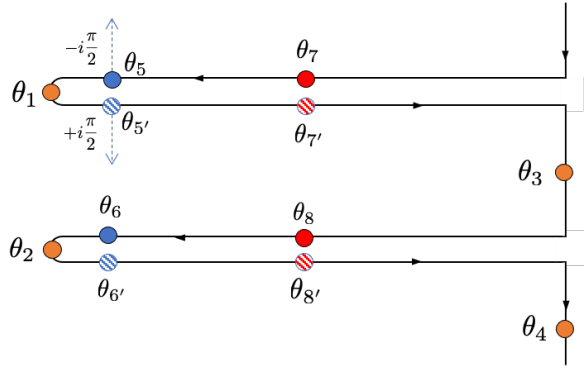

FIG. 2. Double Keldysh contour for the equation (31).

Plugging in the ansatzé for each OTOC on both sides in (31), we can schematically write

$$\frac{e^{\kappa(\mathbf{p})\frac{(t_1+t_2-t_3-t_4)}{2}}}{C(\mathbf{p})}\Upsilon_\mathbf{p}^R(t_{12},\mathbf{k})\Upsilon_\mathbf{p}^A(t_{34},\mathbf{k}') = N\frac{2\cos\frac{\kappa(\mathbf{p})\pi}{2}}{C^2(\mathbf{p})}e^{\kappa(\mathbf{p})\frac{(t_1+t_2-t_3-t_4)}{2}}$$
$$\times \int_{t_5,t_6,t_7,t_8;\mathbf{q},\mathbf{q}'} e^{-\kappa(\mathbf{p})\frac{(t_5+t_6)}{2}}\Upsilon_\mathbf{p}^R(t_{12},\mathbf{k})\Upsilon_\mathbf{p}^A(t_{56},\mathbf{q})\cdot \text{BOX}(t_5,t_6,t_7,t_8;\mathbf{q},\mathbf{q}')\cdot e^{\kappa(\mathbf{p})\frac{(t_7+t_8)}{2}}\Upsilon_\mathbf{p}^R(t_{78},\mathbf{q}')\Upsilon_\mathbf{p}^A(t_{34},\mathbf{k}'),$$

where the dot product is a notation for the integration over the intermediate times and momentum. The factor of $N$ comes from the definition of the OTOC and needs to be compensated since we have two of them on the RHS. Simplifying the equation, we obtain

$$C(\mathbf{p}) = 2N\cos\frac{\kappa(\mathbf{p})\pi}{2}\int_{t_5,t_6,t_7,t_8;\mathbf{q},\mathbf{q}'} e^{\kappa(\mathbf{p})(t_7+t_8-t_5-t_6)/2}\Upsilon_\mathbf{p}^A(t_{56},\mathbf{q})\cdot \text{BOX}(t_5,t_6,t_7,t_8;\mathbf{q},\mathbf{q}')\cdot \Upsilon_\mathbf{p}^R(t_{78},\mathbf{q}').$$

We are interested in the fact that the momentum dependent coefficient $C(\mathbf{p})$ is proportional to

$$\frac{C(\mathbf{p})}{N} \sim \cos\frac{\kappa(\mathbf{p})\pi}{2}, \tag{32}$$

which becomes zero when $\kappa(\mathbf{p}) = 1$. In our case $\lambda_L(\mathbf{p}) = 2\pi T\kappa(\mathbf{p})$, which means that $\lambda_L = 2\pi T$ corresponds to a pole in the regular OTOC (28).

## III. VARYING THE DYNAMICAL CRITICAL EXPONENT

In this section we find a generalized eigenvalue equation when the dynamical critical exponent $2 < z \leq 3$. The retarded boson Green's function and boson Wightman function have the following forms:

$$D^R(\mathbf{q},\Omega \neq 0) = \frac{1}{|q_y|^{z-1} + ic_b\frac{\Omega}{|q_y|}}, \tag{33}$$

$$D^W(\mathbf{q},\Omega) = \frac{c_b\Omega|q_y|}{\sinh\frac{\beta\Omega}{2}(|q_y|^{2z} + c_b^2\Omega^2)}, \tag{34}$$

where $c_b = g^2/8\pi$ as before. The fermion self energy changes to

$$\Sigma(i\omega_n) = ig^2\frac{T}{2}\sum_{\Omega_m \neq 0}\int_{q_y}\frac{\text{sgn}(\omega_n+\Omega_m)}{|q_y|^{z-1} + \frac{g^2}{8\pi}\frac{|\Omega_m|}{|q_y|}} \tag{35}$$
$$= 2^{2-\frac{6}{z}}g^{\frac{4}{z}}T\pi^{1-\frac{2}{z}}\frac{i}{z\sin\frac{2\pi}{z}}\sum_{\Omega_m \neq 0}\Omega_m^{\frac{2}{z}-1}\text{sgn}(\omega_n+\Omega_m)$$
$$= i\text{sgn}(\omega_n)2^{2-\frac{4}{z}}g^{\frac{4}{z}}T^{\frac{2}{z}}\frac{1}{z\sin\frac{2\pi}{z}}H_{1-\frac{2}{z}}\left(\frac{|\omega_n|-\pi T}{2\pi T}\right).$$

In a simpler form, the self energy is

$$\Sigma(i\omega_n) = i\,\text{sgn}(\omega_n) c_{f,z} T^{\frac{2}{z}} H_{1-\frac{2}{z}}\left(\frac{|\omega_n| - \pi T}{2\pi T}\right), \tag{36}$$

$$c_{f,z} = 2^{2-\frac{4}{z}} g^{\frac{4}{z}} \frac{1}{z \sin\left(\frac{2\pi}{z}\right)}, \tag{37}$$

and the fermion Green's function reads

$$G(\mathbf{k}, i\omega_n) = \frac{1}{k_x + k_y^2 - ic_{f,z}\text{sgn}(\omega_n) T^{2/z} H_{1-\frac{2}{z}}\left(\frac{|\omega_n| - \pi T}{2\pi T}\right) - i\,\text{sgn}(\omega_n)\mu(T)}, \tag{38}$$

$$\mu(T) = \frac{g^2 T}{2z \sin\left(\frac{2\pi}{z}\right) m^{2-\frac{4}{z}}}, \tag{39}$$

where $\mu(T)$, as before, is a term generated by a finite but small boson mass $m$. The derivation of the eigenvalue equation is the same as before, and it can be written as

$$\left[T^{2/z} c_{f,z}\left(H_{1-2/z}\left(\frac{-ik_0 - \pi T}{2\pi T}\right) + H_{1-2/z}\left(\frac{-i(\omega - k_0) - \pi T}{2\pi T}\right)\right) + 2\mu(T) - p_x\right]\tilde{\mathcal{C}}(k_0, \omega)$$
$$= g^2 \int \frac{dk_0' dk_y'}{(2\pi)^2} \frac{c_b(k_0 - k_0')|k_y'|}{(|k_y'|^z + m^2)^2 + c_b^2(k_0 - k_0')^2} \frac{\tilde{\mathcal{C}}(k_0', \omega)}{\sinh \frac{k_0 - k_0'}{2T}}$$
$$- \frac{ig^4}{8z \sin\left(\frac{2\pi}{z}\right) c_b^{2-\frac{2}{z}}} \int \frac{dk_0' dk_{01}}{(2\pi)^2} \frac{(-ik_{01})^{\frac{2}{z}-1} - (i(k_{01} - \omega))^{\frac{2}{z}-1}}{2k_{01} - \omega} \frac{\tilde{\mathcal{C}}(k_0', \omega)}{\cosh \frac{k_0 - k_{01}}{2T} \cosh \frac{k_0' - k_{01}}{2T}}. \tag{40}$$

Checking this for $z = 3$, we obtain the result of the previous section. The numerical approach to solve this equation is the same as in the previous section; we vary the external momentum $p_x$ on the imaginary axis and looking for a value of $k_0$ at which the equation has a solution. We present the result of $\lambda_L(|p_x|)/T$ in the main text, and show that any theory with the dynamical critical exponent in a region $2 < z \leq 3$ is maximally chaotic.

### A. Quasiparticles

Here, we would like to explore a regime where quasiparticles appear, specifically when the dynamical critical exponent $z$ is in the region $1 < z < 2$. We explicitly derive and solve the eigenvalue equation, and find and compare the saddle point and pole values. We show that the saddle point gives dominant contribution to the OTOC, and the theories are therefore not maximally chaotic as per the criteria of Ref. [6].

We first compute the fermion self energy with the dynamical critical exponent in the region $1 < z < 2$. As compared to the previous case, we set now the boson mass to zero, but have to include a finite UV cutoff $\Lambda$ when integrating over $k_y$. The fermion self energy reads

$$\Sigma(i\omega_n) = ig^2 \frac{T}{2} \sum_{\Omega_m} \int_{-\Lambda}^{\Lambda} \frac{dq_y}{2\pi} \frac{|q_y|\text{sgn}(\omega_n + \Omega_m)}{|q_y|^z + c_b|\Omega_m|} \tag{41}$$
$$= ig^2 \frac{T}{2\pi} \sum_{\Omega_m} \text{sgn}(\omega_n + \Omega_m) \left[\frac{\Lambda^{2-z}}{2-z} + \frac{\pi c_b^{\frac{2}{z}-1}|\Omega_m|^{\frac{2}{z}-1}}{z \sin\left(\frac{2\pi}{z}\right)}\right]$$
$$= i\omega_n \frac{g^2}{2\pi^2} \frac{\Lambda^{2-z}}{2-z} + ig^2 T^{\frac{2}{z}} \text{sgn}(\omega_n) \frac{(2\pi c_b)^{\frac{2}{z}-1}}{z \sin\left(\frac{2\pi}{z}\right)} H_{1-\frac{2}{z}}\left(\frac{|\omega_n| - \pi T}{2\pi T}\right),$$

where $\Lambda$ is assumed to be large. The retarded self energy is therefore

$$\Sigma^R(\omega) = \omega \frac{g^2}{2\pi^2} \frac{\Lambda^{2-z}}{2-z} + ig^2 T^{\frac{2}{z}} \frac{(2\pi c_b)^{\frac{2}{z}-1}}{z \sin\left(\frac{2\pi}{z}\right)} H_{1-\frac{2}{z}}\left(\frac{-i\omega - \pi T}{2\pi T}\right). \tag{42}$$

The diagonal term in the eigenvalue equation reads

$$\tilde{f}_0(k_0, \omega) = \int \frac{dk_x}{2\pi} G^R(k) G^{R*}(k - \omega) = \frac{i}{\Sigma^R(k_0) - \Sigma^{R*}(k_0 - \omega) + \omega}, \tag{43}$$

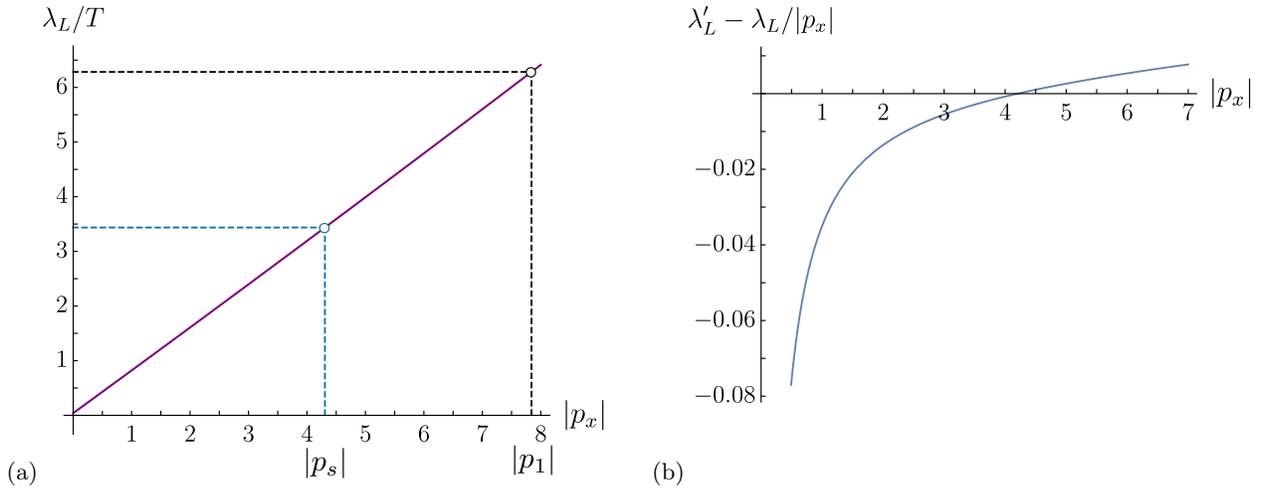

FIG. 3. Analysis of the eigenvalues numerically obtained from (46) for $z = 1.5$, $g = 0.5$ and $\Lambda = 50$. (a) Lyapunov exponent as a function of the external momentum $ip_x$ on the imaginary axis. The saddle point $|p_s|$ gives the dominant contribution as $|p_s| < |p_1|$ [6]. In particular, we obtained $|p_s| = 4.28$ and $|p_1| = 7.84$. The butterfly velocity therefore is $v_B = (d\lambda_L(p_x)/d|p_x|)_{p_x=p_s} = 0.8 v_F$, where $v_F = 1$ is the Fermi velocity. The purple line is a fitting function $\lambda_L = 0.04 + 0.78|p_x| + 0.002|p_x|^2$ of the numerical data. The accuracy of the results is within $\sim 5\%$. (b) Finding the saddle point, which is given by the root of the plotted function.

where we have retained the bare $k_0$ term in the fermion Green's functions, as it is no longer dominated by the self energy unlike the previously considered $2 < z \leq 3$ case. Combining (42) and (43), we obtain

$$\tilde{f}_0(k_0, \omega) = \frac{1}{T^{2/z} c_{f,z} \left( H_{1-2/z} \left( \frac{-ik_0 - \pi T}{2\pi T} \right) + H_{1-2/z} \left( \frac{-i(\omega - k_0) - \pi T}{2\pi T} \right) \right) - i\omega (1 + \frac{g^2}{2\pi^2} \frac{\Lambda^{2-z}}{2-z})}. \tag{44}$$

The $K_1$ term is simply

$$K_1 \tilde{\mathcal{C}} = g^2 \int \frac{dk_0' dk_y'}{(2\pi)^2} \frac{c_b |k_y'| (k_0 - k_0')}{|k_y'|^{2z} + c_b^2 (k_0 - k_0')^2} \frac{\tilde{\mathcal{C}}(k_0', \omega)}{\sinh \frac{k_0 - k_0'}{2T}} = \frac{g^2 c_b^{\frac{2}{z}-1}}{2z \sin \frac{\pi}{z}} \int \frac{dk_0'}{2\pi} \frac{(k_0 - k_0')^{\frac{2}{z}-1}}{\sinh \frac{k_0 - k_0'}{2T}} \tilde{\mathcal{C}}(k_0', \omega), \tag{45}$$

and we note that it is free of IR divergences. We also note that the $K_2$ term is the same as in the previous case (40) for $2 < z \leq 3$, and the eigenvalue equation becomes

$$\left[ T^{2/z} c_{f,z} \left( H_{1-2/z} \left( \frac{-ik_0 - \pi T}{2\pi T} \right) + H_{1-2/z} \left( \frac{-i(\omega - k_0) - \pi T}{2\pi T} \right) \right) - i\omega \left( 1 + \frac{g^2}{2\pi^2} \frac{\Lambda^{2-z}}{2-z} \right) - p_x \right] \tilde{\mathcal{C}}(k_0, \omega)$$
$$= \frac{g^2 c_b^{\frac{2}{z}-1}}{2z \sin \frac{\pi}{z}} \int \frac{dk_0'}{2\pi} \left[ \frac{(k_0 - k_0')^{\frac{2}{z}-1}}{\sinh \frac{k_0 - k_0'}{2T}} - \frac{ig^2}{8 \cos\left(\frac{\pi}{z}\right) c_b} \int \frac{dk_{01}}{2\pi} \frac{(-ik_{01})^{\frac{2}{z}-1} - (i(k_{01} - \omega))^{\frac{2}{z}-1}}{(2k_{01} - \omega) \cosh \frac{k_0 - k_{01}}{2T} \cosh \frac{k_0' - k_{01}}{2T}} \right] \tilde{\mathcal{C}}(k_0', \omega), \tag{46}$$

where the constants $c_{f,z}$ and $c_b$ are defined above.

We solve the equation for parameters $z = 1.5$, $g = 0.5$, $T = 1$, and the UV cutoff $\Lambda = 50$. We find that the ballistic growth is present with the Lyapunov exponent $\lambda_L = 3.42$ and butterfly velocity $v_B = 0.8 v_F$. Since the solution is numerical, we obtain the result with some error, that we estimate to be $\sim 5\%$. We show the behavior of eigenvalues upon changing the external momentum on the imaginary axis $|p_x|$ in Fig. 3. We find that the saddle point value is $|p_s| = 4.28$, and that the pole is at $|p_1| = 7.84$ which is much greater than the saddle point. Therefore, there is no maximal chaos according to the criteria set by Ref. [6].

We also solve the eigenvalue equation for several other values of the dynamical critical exponent in the range $1 < z < 2$, where quasiparticles are present. As in the main text, we compute the difference between the velocities $v_1$ and instantaneous slopes $v_*$ at $p_x = p_1$. We show the result in Fig. 4. Within our numerical accuracy ($\sim 5\%$), we find that the instantaneous slope $v_*$ is always larger than the velocity $v_1$ at the pole $|p_1|$. Since $\lambda_L(|p_x|)$ is a positive, monotonically increasing, and convex function, this implies $|p_s| < |p_1|$. Therefore, following the criteria set by Ref. [6], we can argue that for $z < 2$, the theories are not maximally chaotic.

We also note that the leading behavior of the eigenvalues is controlled by the linear in frequency behavior in the diagonal term of the equation (46). Therefore, to first order, we see that $\lambda_L \sim T^{2/z} \ll 2\pi T$, and we can also estimate

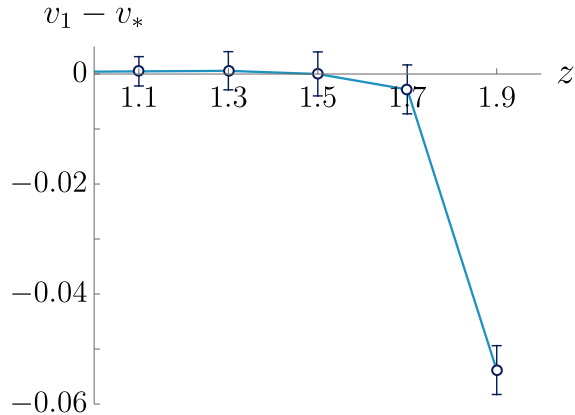

FIG. 4. Difference between $v_1$ and the instantaneous slopes $v_*$ at $p_x = p_1$, for several dynamical critical exponents $z$ in the region $1 < z < 2$. This difference is negative within numerical accuracy, and therefore there is no maximal chaos when $z < 2$. This is in contrast to the $2 < z \leq 3$ case, where the difference is always positive (Fig. 5 of the main text)

the butterfly velocity as

$$v_B \approx \frac{1}{1 + \frac{g^2}{2\pi^2}\frac{\Lambda^{2-z}}{2-z}} v_F, \qquad (47)$$

where $v_F = 1$ is the Fermi velocity. For the chosen parameters, this equation approximates $v_B \approx 0.85\, v_F$, which is close to the actual numerical solution we obtained above. We therefore argue that, within the class of problems we are interested in, any theory with the dynamical critical exponent in the region $1 < z < 2$, *i.e.*, one that has quasiparticles, is not maximally chaotic.

[1] D. Chowdhury and B. Swingle, Onset of many-body chaos in the $O(N)$ model, Phys. Rev. D **96**, 065005 (2017).
[2] A. A. Patel and S. Sachdev, Quantum chaos on a critical Fermi surface, Proc. Nat. Acad. Sci. **114**, 1844 (2017), arXiv:1611.00003 [cond-mat.str-el].
[3] J. Steinberg and B. Swingle, Thermalization and chaos in QED$_3$, Phys. Rev. D **99**, 076007 (2019).
[4] I. Esterlis, H. Guo, A. A. Patel, and S. Sachdev, Large $N$ theory of critical Fermi surfaces, Phys. Rev. B **103**, 235129 (2021), arXiv:2103.08615 [cond-mat.str-el].
[5] A. Kitaev and S. J. Suh, The soft mode in the Sachdev-Ye-Kitaev model and its gravity dual, JHEP **05**, 183, arXiv:1711.08467 [hep-th].
[6] Y. Gu and A. Kitaev, On the relation between the magnitude and exponent of OTOCs, JHEP **02**, 075, arXiv:1812.00120 [hep-th].
[7] Y. Gu, A. Kitaev, and P. Zhang, A two-way approach to out-of-time-order correlators (2021), arXiv:2111.12007 [hep-th].



\end{document}